\documentclass[aps,prx,twocolumn,amsmath,amsfonts,amssymb,floatfix,superscriptaddress,nofootinbib]{revtex4}
\usepackage{epsfig,amsmath,amssymb,graphics,color,calc}
\usepackage{mathtools}
\pdfoutput=1

\allowdisplaybreaks

\DeclarePairedDelimiter\bra{\langle}{\rvert}
\DeclarePairedDelimiter\ket{\lvert}{\rangle}
\DeclarePairedDelimiterX\braket[2]{\langle}{\rangle}{#1 \delimsize\vert #2}

\begin{document}

\title{Critical properties of the Anderson localization  transition and the high dimensional limit}

\author{E. Tarquini\textsuperscript{1,2,4}, G. Biroli\textsuperscript{2,3}, and M. Tarzia}

\affiliation{\mbox{LPTMC,~CNRS-UMR 7600,~Sorbonne Universit\'e,
4 place Jussieu, 75252 Paris c\'edex 05, France}\\\textsuperscript{2} \mbox{Institut de physique th\'eorique, Universit\'e Paris Saclay, CEA, CNRS, F-91191 Gif-sur-Yvette, France}\\ \textsuperscript{3}Laboratoire de
  Physique Statistique, Ecole Normale Sup\'erieure, PSL Research
  University, 24 rue Lhomond, 75005 Paris, France.
  \\ \textsuperscript{4} \mbox{Universit\'e Paris-Sud, 91405-ORSAY, France}}

\begin{abstract}
In this paper we present a thorough study of transport, spectral and wave-function properties at the Anderson localization critical point in spatial dimensions $d = 3$, $4$, $5$, $6$. Our aim is to analyze the dimensional dependence and to asses the role of the $d\rightarrow \infty$ limit provided by Bethe lattices and tree-like structures. 
Our results  strongly suggest that the upper critical dimension of Anderson
localization is infinite.  Furthermore, we find that the $d_U=\infty$ is a much better starting point 
compared to $d_L=2$ to describe even three dimensional systems. 
We find that critical properties and finite size scaling behavior approach by increasing $d$
the ones found for Bethe lattices: the critical state becomes an insulator characterized by Poisson statistics 
and corrections to the thermodynamics limit become logarithmic in $N$. 
In the conclusion, we present physical consequences of our results,  propose connections with the 
non-ergodic delocalised phase suggested for the Anderson model on infinite dimensional lattices and 
discuss perspectives for future 
research studies.  
\end{abstract}


\maketitle

\section{Introduction}



Anderson localization (AL) is one of the most fundamental quantum phenomena. 
A system of non-interacting disordered electrons can be driven  (e.g.,  by  increasing  the disorder strength or the energy) 
through a transition between a metallic (delocalized) phase and insulating (localized) phase, 
where diffusive transport is completely suppressed due to quantum interference~\cite{anderson}.
After more than half century of research~\cite{fiftylocalization}, the subject is still very much alive as
proved by recent experimental observation of AL in $1d$~\cite{aspect} and $3d$~\cite{chabe} atomic gases
and for classical sound elastic waves in $3d$~\cite{localizationelastic}.


The properties of AL in low dimensional systems are by now very well established and understood.
As predicted by the scaling theory of localization~\cite{scaling}, all states are localized in $1d$~\cite{mott} and $2d$
(for system with orthogonal symmetry) by an infinitesimal amount of disorder. 
In fact, $d_L=2$ is the lower critical dimension of the problem,
where the so-called ``weak localization'' takes place~\cite{weak}.

During the last 40 years, a field theoretical approach~\cite{NLsM} based on the 
replicated Non-Linear $\sigma$-Model (NL$\sigma$M) has been developed, 
and 
a perturbative $\epsilon$ expansion in  $d=2+\epsilon$ dimensions has been 
pushed up to five-loops~\cite{5-loops}.
These advances culminated in a functional (perturbative) renormalization group analysis~\cite{ludwig} of the NL$\sigma$M, 
which allowed to compute the multifractal spectra of wave-function amplitudes at the AL critical point
in $d=2+\epsilon$.


Nonetheless, despite about 60 years of intense research, there is still (almost) no available analytical approach 
for AL away from the low-dimensional limit and much less is known in higher dimensions.
The main reasons for that are: 
\begin{itemize}
\item[(a)] The absence of small parameter: The critical disorder is of the 
same order (or even larger) than the bandwidth already in three dimensional systems.
\item[(b)] The fact that AL is not associated to a conventional spontaneous symmetry breaking. 
Indeed, the order parameter which naturally arises in the field theoretical
description is a function: the probability distribution
of the local density of states (DOS) which develops heavy tails in the insulating phase
due to very large and rare resonances. The average DOS instead does not show any sign of
discontinuity at the transition. 
\end{itemize}
These unconventional properties represent a challenge for analytical approaches.
As a consequence numerical methods are still at the core of the advances in this topic~\cite{review-numerics}.


AL in three dimensions was analyzed by many authors using numerical techniques
for increasing system size, 
with the use of various scaling analysis, and of different observables related both
to transport properties~\cite{numerics3d,best3d1} and to the statistics of energy levels~\cite{levelstatistics3d} and wave-functions
coefficients~\cite{best3d2,multi3d}.
In Ref.~\cite{EW3d} the phase diagram in the energy-disorder plane was also calculated.
For the model described in the next section (spinless electron in a uniformly distributed
disordered potential) and for $E=0$ (middle of the band) a localization transition is found at a critical value of
the disorder $W_c \simeq 16.5$, separating a metallic phase, where wave-functions are 
extended over the whole volume, from an insulating phase, where wave-functions
are exponentially localized around some particular sites.
The quantity $\overline{\Upsilon_2} = \overline{\sum_i |\braket{n}{i}|^4}$, called the inverse participation ratio 
(IPR)---averaged over the disorder and over
all eigenstates $\ket{n}$ around $E=0$---is
often used to distinguish between 
these two regimes as $\overline{\Upsilon_2} \sim C/L^d$ 
in the extended phase and
stays of $O(1)$ in the localized phase.
Diffusion is completely suppressed in the insulating regime and the
conductivity $\sigma$ vanishes in the thermodynamic limit, while 
it stays finite in the metallic phase. The localization length, measuring the spatial extent over which 
wave-functions are localized, diverges at the transition coming from the insulating phase. 
At present, the most precise numerical estimate of the critical exponent $\nu$ describing this
divergence in $3d$---for systems with orthogonal
symmetry---is $\nu = 1.58 \pm 0.01$~\cite{best3d1,best3d2}.

AL had a very strong impact also on Random Matrix Theory (RMT). 
As a matter of fact, in the delocalized phase the level statistics on the scale of the mean-level spacing 
is expected to be described by RMT and generally corresponds
to the Gaussian Orthogonal Ensemble (GOE), whereas instead
in the localized phase is determined by Poisson statistics because wave-functions close in energy are exponentially
localized on very distant sites and hence do not overlap; thus, contrary to the GOE case, there is no level-repulsion and eigen-energies
are distributed similarly to random points thrown on a line. These ideas 
have been confirmed by numerical simulations in $3d$~\cite{levelstatistics3d}.

Right--and only--at the critical point, level statistics is neither GOE nor Poisson~\cite{LScrit} (it is
instead characterized by a universal distribution which depends on the dimensionality) and 
wave-function amplitudes show a multifractal spectrum~\cite{multi3d}---the 
critical eigenstates being neither extended nor localized
reveal large fluctuations of wave-function amplitudes at all
length scales.

Few recent accurate results are also available in $4d$ and $5d$~\cite{numerics4d5d}, based on the study of
transport properties only. 
However, 
there are very few results on level statistics above dimension three~\cite{garcia} and no 
exact results for transport properties for $d>5$~\cite{scardicchio-highd}. As we will discuss in the following, the reason for that 
is that running times of numerical algorithms increase very rapidly with the size of the system 
(more precisely, as $L^{3d}$ for exact diagonalization (ED) and as $L^{3d-2}$ for transfer matrix (TM) techniques).
This sets a very severe limitation on the system sizes which can be simulated
as dimensionality is increased.


For these reasons, some basic questions of AL 
remain unanswered or debated.
For instance, the existence of an upper critical dimension $d_U$ is still an issue.
Although several observations seem to indicate that $d_U$ 
might be infinite~\cite{garcia,duinfinite}, different propositions corresponding
to $d_U=4$, $6$, and $8$ have been put forward~\cite{otherdu,vollhardt}.

Another important and highly debated aspect is the relation with the infinite $d$ 
limit, corresponding to AL on tree-like structures~\cite{abou} 
and to other random matrix models with long-range hopping~\cite{levy}.
On the one hand, these models allow for an exact solution, making it possible
to establish the transition point and the corresponding critical behavior~\cite{levy,infinitedexact,SUSY-tree}.
On the other hand, however, the properties of the delocalized phase are very unusual, 
since they are affected by 
dramatic---and somehow unexpected---finite-size effects (FSE) even very far from the critical point,
which produce a strong non-ergodic behavior in a crossover region 
where the correlation volume
is larger than the accessible system sizes~\cite{levy,noi,scardicchio,altshuler,mirlin1,mirlin2,lemarie}. 
This makes the finite-size analysis of numerical 
data highly non-trivial~\cite{mirlin1,lemarie}, and has been
interpreted by some authors~\cite{noi,scardicchio,altshuler} in terms of the existence of 
a new intermediate delocalized but non-ergodic 
phase---which might be characterized by non-universal level statistic, anomalous scaling exponents of
the IPR, and multifractality~\cite{kravtsov}---in
a broad interval of disorder strength between the metallic (fully ergodic) phase and the insulating one.

This possibility is clearly very intriguing (although it appears to be in conflict with the
analytical predictions of the SUSY formalism~\cite{SUSY-tree}), especially due to its relationship with Many-Body
localization~\cite{BAA}, a fascinating new kind of phase transition between a
low temperature non-ergodic phase---a purely quantum
glass---and a high temperature ergodic phase.
Theoretical work strongly suggests that this phenomenon takes place for several
disordered isolated interacting quantum systems, in particular
disordered electrons~\cite{BAA} (it was also independently investigated
in~\cite{wolynes} to explain the quantum ergodicity
transition of complex molecules).
MBL can be pictorially interpreted 
as localization in
the Fock space of Slater determinants, which play the
role of lattice sites in a disordered (single-particle) Anderson tight-binding model.
A paradigmatic representation of this transition~\cite{A97,BAA,jacquod,wolynes,scardicchioMB}
is indeed AL on a very high dimensional lattice, 
which for spinless electrons consists in an $N$-dimensional hyper-cube of $2^N$ sites. 



All the open questions presented above motivated us to thoroughly analyse AL in high spatial dimensions.  
In the following we present a detailed study of the critical properties of AL in dimensions from $3$ to $6$ based on ``exact'' numerical methods (ED and TM techniques) and on an approximate
Strong Disorder Renormalization Group (SDRG) approach~\cite{sdrg,sdrg1d}.
We focus on both the statistics of energy levels and wave-functions coefficients and
on transport properties. 
Our aim is to shed new light on the critical properties of AL 
and provide new insights to develop alternative analytical approaches to tackle this problem.


Our results support the idea that the upper critical dimension of AL is infinite.
For instance, the critical exponent $\nu$ smoothly evolves from $\nu \to \infty$ in $d=2$ to the value
$\nu = 1/2$ in $d \to \infty$ predicted by the SUSY approach~\cite{SUSYdinf}, showing no sign of saturation.
Moreover,  we find that the infinite dimensional limit is a very good 
quantitative and qualitative starting point to describe AL even down to three dimensions. 
Expansions around the lower critical dimension, $d_L=2$, instead give 
poorer results (even up to five-loops).  
The higher is the dimension the more AL is well described by a strong disorder limit, 
as signalled by the fact that the critical values of all observables smoothly approach the ones of the localized phase as the 
dimensionality is increased---in $d \to \infty$ the critical states seem to correspond to an insulator, for which the statistics
of energy levels is of Poisson type, 
and the multifractal spectrum of wave-functions amplitudes takes its
strongest possible form.
Another strong indication of this fact is that the SDRG approach 
gives very accurate results in estimating the critical parameters in all dimensions $d \ge 3$.

We also show that FSE become anomalously strong as $d$ is increased. 
When $d$ gets large the scaling variable controlling finite size scaling (FSS)
is $|W-W_c| L^{1/\nu}$ and the leading corrections to FSS turn out to be proportional to 
$L^y$. Both $\nu$ and $y$ depend weakly on the dimensions and tend to a constant
when $d\rightarrow \infty$: $\nu \rightarrow 1/2$ and $y$ stays roughly constant and close to $-1$.
When re-expressed in terms of the systems size $N=L^{1/d}$ these results suggest 
that corrections become logarithmic-like in $N$ in the $d \to \infty$ limit. 
This behavior is drastically different from the one observed in conventional 
phase transitions, for which it exists an upper critical dimension $d_U$ such that 
for $d>d_U$ finite size effects are governed by the scaling variable $|T-T_c| N^{1/\nu d_U}$ with corrections 
of the order of $N^{y^\prime}$ (with some negative exponent $y^\prime$ independent of $d$)

The paper is organized as follows:
In Sec.~\ref{sec:model} we introduce the model and some basic definitions. In Sec.~\ref{sec:numerics} we
present our numerical results based on exact diagonalization (ED) and transfer matrix (TM) methods 
for dimensions from $3$ to $6$. In Sec.~\ref{sec:SDRG} we discuss the SDRG approach, focusing especially
on the properties of the flow close to criticality.
In Sec.~\ref{sec:results} we give a brief summary of the results found and discuss the
their possible implications on the unusual properties of the delocalized phase
observed 
in the Anderson model on tree-like structures,
which can be interpreted in terms of the extreme ``quasi-localized'' character of the AL critical point in $d 
\to \infty$, and of anomalously
strong FSE. Finally, in Sec.~\ref{sec:conclusions} we present some concluding remarks and 
perspectives for future work.

\section{The model} \label{sec:model}
The model we focus on consists in non-interacting spinless electrons in a
disordered potential: 
\begin{equation} \label{eq:H}
{\cal H} = - t \sum_{\langle i,j \rangle} \left( c_i^{\dagger} c_j
+ c_j^{\dagger} c_i \right ) - \sum_{i=1}^N \epsilon_i c_i^\dagger c_i \, ,
\end{equation}
where the second sum runs over all $N = L^d$ sites, and
the first sum runs over all $d L^d$ links of nearest neighbors sites of the
$d$-dimensional hyper-cubic lattice; 
$c_i^\dagger$, $c_i$ are fermionic creation and annihilation operators, and
$t$ is the hopping kinetic energy scale, which we take
equal to $1$ throughout.
The on-site energies $\epsilon_i$ are i.i.d. random variables uniformly
distributed in the interval $[-W/2,W/2]$:
\begin{equation} \label{eq:peps}
p(\epsilon) = \frac{1}{W} \, \theta \! \left ( \frac{W}{2} - | \epsilon | \right) \, ,
\end{equation}
$W$ being the disorder strength.
The model~(\ref{eq:H}) has time reversal (and spin rotation) symmetry (also called {\it orthogonal} symmetry 
in the context of RMT).
The common belief, supported by the scaling theory of localization~\cite{scaling} is that the 
transition is universal, i.e., it does not depend on microscopic details of the model such as the probability distribution
of the on-site energies. However, it depends on the dimension and on the physical symmetry of ${\cal H}$. 

\begin{figure}
 \includegraphics[angle=0,width=0.44\textwidth]{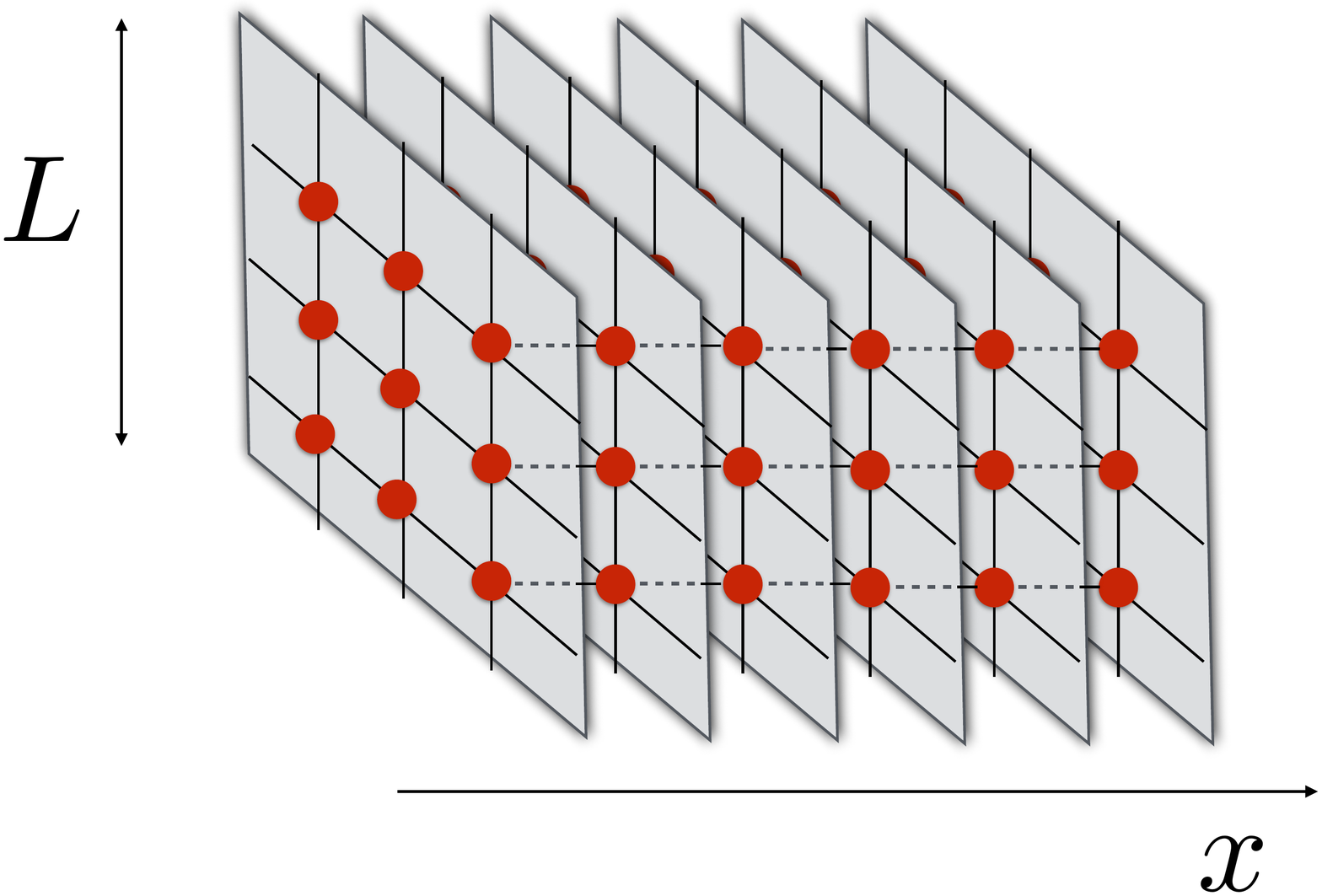}
 \caption{\label{fig:quasi-1d}
Sketch of the quasi-one dimensional bar along the $x$ direction of cross section
$L^{(d-1)}$.}
 \end{figure}

In terms of RMT, the model~(\ref{eq:H}) can be thought as a sum of two matrices, 
${\cal H} = {\cal C}^{(d)} + {\cal E}$
(i.e., a Schr\"odiger operator with random on-site potential):
${\cal C}^{(d)}$ is the (deterministic) connectivity matrix of the 
$d$-dimensional hyper-cube, 
${\cal C}^{(d)}_{ij} = -t$ if sites $i$ and $j$ are connected
and zero otherwise. ${\cal E}$ is a diagonal random matrix corresponding to the on-site energies,
${\cal E}_{ij} = \epsilon_i \delta_{ij}$.

In the following we will focus only on the middle of the spectrum, $E=0$.

\section{Numerical results in $d=3,\ldots,6$} \label{sec:numerics}

In this section we present our numerical results in dimensions 
from $3$ to $6$ obtained from ED and a TM approach.
We will focus first on transport properties and then
on the statistics of energy gaps and wave-functions amplitudes.

\subsection{Transport properties} \label{sec:lyapunov}

\begin{figure}
 \includegraphics[angle=0,width=0.44\textwidth]{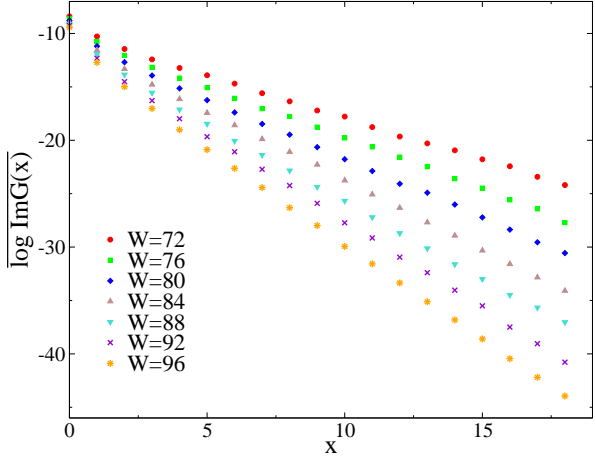}
 \caption{\label{fig:lnImGx}
$\overline{\log \mbox{Im} G (x)}$ as a function of $x$ in $6$ dimensions, for $L=6$ and for several values of
the disorder, showing that $\xi_{1d}$ can be measured from Eq.~(\ref{eq:xi1d})
by linear fitting of the data at large enough $x$.}
 \end{figure}

We consider a very long (length $L_x$) quasi-one dimensional bar of cross-section $L^{d-1}$, as sketched
in fig.~\ref{fig:quasi-1d}.
The system is open along the $x$-direction, while periodic boundary conditions 
are enforced along the transverse 
directions.
Such system, being quasi-$1d$, is always localized at any arbitrarily weak value of the disorder.
The localization of electrons on this bar can be studied using the TM method. 
To this aim, 
we introduce the resolvent matrix, ${\cal G} = [z {\cal I} - {\cal H}]^{-1}$, where $z = E + i \eta$ with $\eta \to 0^+$ being 
the imaginary regulator, and express its matrix
elements in terms of a Gaussian integral over a real auxiliary field: 
\begin{equation} \label{eq:resolvent}
{\cal G}_{lm} = - \frac{i}{Z} \int \prod_{i=1}^N {\rm d} \phi_i  \, \phi_l \phi_m \, 
e^{ S[\phi_i] } \, ,
\end{equation}
where the action is given by:
\begin{equation} \label{eq:action}
\begin{split}
S[\phi_i] & = \frac{i}{2} \sum_{i,j=1}^N \phi_i
\left( z \delta_{ij} - {\cal H}_{ij} \right) \phi_j \\
& = \frac{i}{2} \sum_i \left(E + i \eta + \epsilon_i \right) \phi_i^2 + i \sum_{\langle i,j \rangle} t_{ij} \phi_i \phi_j \, ,
\end{split}
\end{equation}
and the ``partition function'' reads:
\begin{equation} \label{eq:Z}
Z = \int \prod_{i=1}^N {\rm d} \phi_i  \, e^{ S[\phi_i] } \, .
\end{equation}
We set $E=0$ throughout, which corresponds to the band center.
We set a finite positive
value of $\eta$ at $x=0$ and $\eta = 0$ elsewhere 
inside the bar, at $x>0$.
This mimics putting the left boundary of the quasi-$1d$ bar of fig.~\ref{fig:quasi-1d} in contact with a 
thermal bath, and study how dissipation propagates through the sample.
The quasi-$1d$ localization length, $\xi_{1d}$, can be easily measured from
the exponential decay of the typical value of the imaginary part of the       
Green's function, $\exp[ \overline{\log \mbox{Im}{\cal G}(x) }]$, as a function of $x$, averaged over all the 
sites of the $x$-th layer and over several realizations of the disorder:
\begin{equation} \label{eq:xi1d}
\overline{\log \mbox{Im} {\cal G} (x)} \simeq {\rm cst} - \frac{x}{\xi_{1d}} \, .
\end{equation}
Since Eq.~(\ref{eq:action}) is a Gaussian action, 
in order to compute the l.h.s. of Eq.~(\ref{eq:xi1d})
one can---at least formally---integrate over all the sites on a given layer $x$ in Eq.~(\ref{eq:resolvent}), 
yielding an exact recursive relation expressing the 
Green's function on the subsequent layer, $x+1$, in terms of the Green's function on the layer $x$ in absence 
of layer $x+1$ (a kind of cavity equation for the whole layer):
\begin{equation} \label{eq:cavity-quasi1d}
\left[{\cal G} (x+1) \right]^{-1}_{ij} = \epsilon_{x,i} \delta_{ij} + t {\cal C}^{(d-1)}_{ij}
- t^2 {\cal G}_{ij} (x) \, ,
\end{equation}
where the index $i$ runs over all the sites of layer $x$, $\epsilon_{x,i}$ is the random on-site energy on site $i$
belonging to layer $x$, and ${\cal C}^{(d-1)}$ is the connectivity matrix of the transverse layers, i.e., the $(d-1)$-dimensional
hyper-cube. 
This equation can be solved numerically by iteration, starting from the following initial condition at $x=0$:
\begin{equation}
\left[{\cal G} (0) \right]^{-1}_{ij} = \left( \epsilon_{0,i} + i \eta \right ) \delta_{ij} + t {\cal C}^{(d-1)}_{ij} \, .
\end{equation}
In order to do this we need to invert the matrix ${\cal G} (x)$ layer by layer, which can be done 
by LU decomposition.
Since the computer time required to perform this operation is proportional to the third power of the 
total number of sites of the matrix, $L^{3(d-1)}$, the running time of the TM
algorithm scales as $L_x L^{3d-3} \sim L^{3d - 2}$.

\begin{figure}
 \includegraphics[angle=0,width=0.48\textwidth]{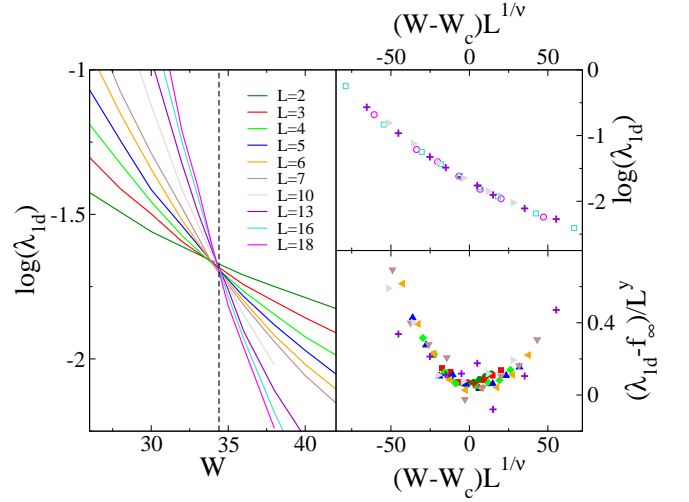}
 \caption{\label{fig:lambda1d-4d}
Left panel: $\lambda_{1d}$ as a function of the disorder $W$ for several
 system sizes $L$ from $2$ to $18$. The vertical
dashed line spots the position of the critical point, $W_c \simeq 34.5$.
Top-right panel: Finite size scaling of the
same data for $L$ from $10$ to $18$, showing data collapse for $\nu \simeq 1.11$.
Bottom-right panel: $\psi f_1 = (\lambda_{1d} - f_\infty)/L^y$ as a function of the
scaling variable $(W-W_c) L^{1/\nu}$ for different sizes $L$ from $2$ to $7$, 
showing data collapse for the same value as before of $W_c$ and $\nu$
and for $y \simeq -1$.}
 \end{figure}

As an example, in fig.~\ref{fig:lnImGx} we plot $\overline{\log \mbox{Im}G (x)}$ as a function of $x$ in $6$ dimensions, 
for $L=6$ and for several values of the disorder $W$, showing that $\xi_{1d}$ can be measured using Eq.~(\ref{eq:xi1d}) 
by linear fitting of the data at large enough $x$.
This is equivalent to the following definition of the quasi-$1d$ localization length via the trasmission 
coefficient~\cite{numerics3d,scaling-channels}:
\begin{displaymath}
\xi_{1d}^{-1} = - \lim_{L_x \to \infty} \frac{1}{2 (L_x+1)} \, \log \, \mbox{Tr} | \bra{0} {\cal G} \ket{L_x} |^2 \, ,
\end{displaymath}
where $\bra{0} {\cal G} \ket{L_x}$ denotes the $L^{d-1}$-dimensional matrix of the resolvent between the
site states in the $0$-th and $L_x$-th slice of the system
(i.e., $\mbox{Tr} | \bra{0} {\cal G} \ket{L_x} |^2$ is the probability for an electron to go from a site on the 
layer $0$ to a site on the layer $L_x$).
One can then work out the asymptotic behavior of $\xi_{1d}$:
In the localized regime one expects that for $L$ large enough $\xi_{1d}$ saturates to the
actual value of the localization length $\xi$ of the $d$-dimensional system.
Conversely, in the extended regime the wave travelling along the bar is evenly spread over
the whole bar. The effective disorder seen by the
wave in each layer 
is thus a statistical average
over the disorder in the layer. One can show that the results of perturbation theory for $1d$ are also valid here, with
the modified disorder $\tilde{W}^2 = W^2 / L^{d-1}$~\cite{scaling-channels}.
As a result, one expects that in the metallic phase $\xi_{1d}$ grows as $L^{d-1}$, 
i.e., the number of (open) channels in the transverse direction. 
(Note that in this case the correlation lenght $\xi$ is related to the resistivity of the
$d$-dimensional system via $\sigma \propto 1/\xi^{d-2}$~\cite{scaling-channels}).

Hence, the good scaling variable is the dimensionless quasi-$1d$ localization length, defined as 
$\lambda_{1d} = \xi_{1d}/L$. This quantity is the inverse of the smallest positive Lyapunov exponent $\gamma$, and 
behaves as:
\begin{displaymath}
\lambda_{1d} \simeq \left \{
\begin{array}{ll}
(L/\xi)^{d-2} \propto \sigma L^{d-2}~~~ & {\rm for}~W<W_c \\
\lambda_c & {\rm for}~W=W_c \\
\xi/L & {\rm for}~W>W_c 
\end{array}
\right .
\end{displaymath}

\begin{figure}
 \includegraphics[angle=0,width=0.48\textwidth]{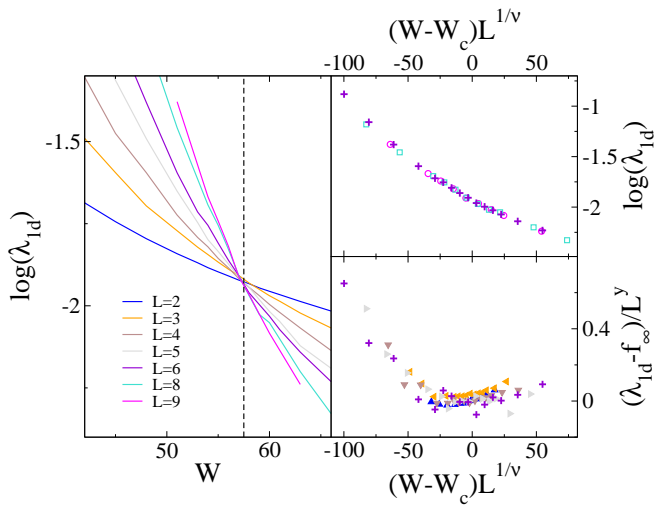}
 \caption{\label{fig:lambda1d-5d}
Left panel: $\lambda_{1d}$ as a function of the disorder $W$ for several
 system sizes $L$ from $2$ to $9$. The vertical
dashed line spots the position of the critical point, $W_c \simeq 57.5$.
Top-right panel: Finite size scaling of the
same data for $L$ from $6$ to $9$, showing data collapse for $\nu \simeq 0.96$.
Bottom-right panel: $\psi f_1 = (\lambda_{1d} - f_\infty)/L^y$ as a function of the
scaling variable $(W-W_c) L^{1/\nu}$ for different sizes $L$ from $2$ to $6$,
showing data collapse for the same value as before of $W_c$ and $\nu$
and for $y \simeq -1.2$.}
 \end{figure}

\begin{figure}
 \includegraphics[angle=0,width=0.48\textwidth]{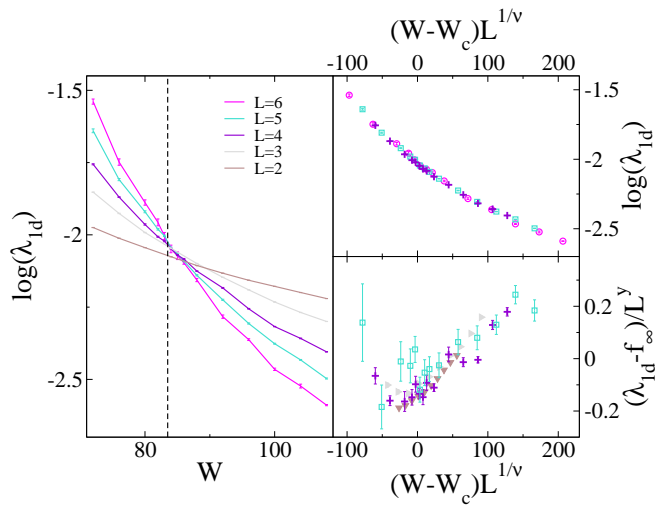}
 \caption{\label{fig:lambda1d-6d}
Left panel: $\lambda_{1d}$ as a function of the disorder $W$ for several
 system sizes $L$ from $2$ to $6$. The vertical
dashed line spots the position of the critical point, $W_c \simeq 83.5$.
Top-right panel: Finite size scaling of the
same data for $L$ equal to $4$, $5$ and $6$, showing data collapse for $\nu \simeq 0.84$.
Bottom-right panel: $\psi f_1 = (\lambda_{1d} - f_\infty)/L^y$ as a function of the
scaling variable $(W-W_c) L^{1/\nu}$ for different sizes $L$ from $2$ to $5$,  
showing data collapse for the same value as before of $W_c$ and $\nu$
and for $y \simeq -1.4$.}
 \end{figure}

The left panels of figs.~\ref{fig:lambda1d-4d}, \ref{fig:lambda1d-5d}, and \ref{fig:lambda1d-6d} show
the behavior of (the log of) the dimensionless quasi-$1d$ localization length $\lambda_{1d}$ 
as a function of $W$ for several system sizes in
dimensions $4$, $5$ and $6$ respectively.
As expected, for small (resp. large) values of the disorder $\lambda_{1d}$ grows (resp. decreases) as $L$ 
is increased; For large enough sizes, the curves corresponding to different $L$ cross at the critical point.
However, the figures show
the presence of systematic FSE due to practical limitations on the system sizes: 
In $4d$ the crossing point shifts 
towards higher values of $W$ by about $2.5 \%$ as $L$ is increased from $2$ to $18$, 
while in $5d$ it moves 
towards lower values of the disorder (again by about $2.5 \%$) when $L$ goes from $2$ to $9$.
FSE become very strong in $6d$, 
where the crossing point shifts systematically 
to lower values of $W$ by about $10 \%$ when $L$ varies from $2$
to $6$.
This gives a first qualitative indication of the fact that, differently from conventional phase transitions, 
FSE for AL get stronger as the dimensionality is increased.

Such finite-size corrections must thus be taken into account in
in order to get accurate
estimations of the critical values of the disorder strength and of the critical exponent.
This can be done considering the presence of irrelevant scaling variables. More precisely, we follow~\cite{best3d1,numerics4d5d} and suppose that
the dependence of $\lambda_{1d}$ 
on $W$ and $L$ can be described in terms of a scaling function:
\begin{equation} \label{eq:two-parameter-scaling}
\lambda_{1d} (W,L) = F \! \left( \! w L^{1/\nu}, \psi L^y \right) \, , 
\end{equation}
where $w=(W-W_c)/W_c$ is the (dimensionless) distance from the critical point, $\nu$ is the critical exponent,
$\psi$ is the leading irrelevant scaling variable, and $y$ is
the smallest (in absolute value) irrelevant critical exponent (consistently,
we should find $y<0$ if $\psi$ is irrelevant).
For finite $L$ there is no phase transition and $F$ is a smooth function
of its arguments. Hence, assuming that the irrelevant scaling variable is not dangerous (and
for $L$ large enough),
one can expand Eq.~(\ref{eq:two-parameter-scaling}) up to first order in $\psi L^y$:
\begin{equation} \label{eq:tps}
\lambda_{1d} (W,L) = f_\infty \! \left( \! w L^{1/\nu} \right) +  \psi L^y f_1 \! \left( \! w L^{1/\nu} \right) \, .
\end{equation}
In order to estimate $W_c$, $\nu$ and $y$ we then proceed in the following way:
\begin{itemize}
\item[(1)] Since FSE are negligible for $L$ large enough, we suppose that one can obtain an
approximate evaluation of
the function $f_\infty (x)$ by performing a cubic fit of the numerical data for the largest available
system sizes (in practice  we use $L = 18$ and $16$ in $d=4$, $L = 9$ and $8$ in $d=5$, and $L = 6$ in $d=6$).
Note that the validity of this assumption must be verified {\it a posteriori}, since it depends on
the value the irrelevant exponent $y$, on $L_{\rm max}$, and on the form of the scaling function $f_1$.
\item[(2)] We plot the difference between the numerical data for $L<L_{\rm max}$ and 
the function $f_\infty$ estimated in step (1), divided by $L^y$, 
as a function of the scaling variable $(W-W_c) L^{1/\nu}$. 
We determine the values of $W_c$, $\nu$ and $y$ that give the best data collapse
of the curves corresponding to different values of $L$ 
(see bottom-right panels of figs.~\ref{fig:lambda1d-4d}, \ref{fig:lambda1d-5d}, and \ref{fig:lambda1d-6d}), 
yielding an approximate estimation of ($\psi$ times) the scaling function $f_1$ (which can 
also be approximated by a cubic fit).
\item[(3)] We plot $\lambda_{1d}$ as a function of $(W-W_c) L^{1/\nu}$ for the largest
sizes only, checking that our estimation of the critical parameters give a good data collapse
(see top-right panels of figs.~\ref{fig:lambda1d-4d}, \ref{fig:lambda1d-5d}, and \ref{fig:lambda1d-6d}).
\item[(4)] Having estimated the scaling function $\psi f_1$ and the critical parameters $W_c$, $\nu$, and $y$ in the 
previous steps, we can iteratively improve the estimation
of $f_\infty$ obtained in step (1) by performing a cubic fit of $\lambda_{1d} (W, L_{\rm max}) - \psi L_{\rm max}^y f_1 (W, L_{\rm max})$,
which takes into account finite-size corrections also for the largest system size in a self-consistent way.
One can then repeat the whole process until it converges. 
\end{itemize}
This analysis yields the following values for the critical parameters:
\begin{equation} \label{eq:critical}
\begin{array}{c|c|c}
d=4 & d=5 & d=6 \\
\hline
W_c = 34.5 \pm 0.2 & W_c = 57.5 \pm 0.2 & W_c = 83.5 \pm 0.4 \\
\nu = 1.11 \pm 0.05 & \nu = 0.96 \pm 0.06 & \nu = 0.84 \pm 0.07 \\
y = -1 \pm 0.1 & y = -1.2 \pm 0.1 & y = -1.4 \pm 0.2
\end{array}
\end{equation}
The results in $4d$ and $5d$ are in excellent agreement with 
the recent accurate estimations of~\cite{numerics4d5d}, while our analysis provides the first direct calculation  
of the critical parameters for AL in six dimensions.\footnote{Note, however, that in order for the assumption in (1) to be correct, 
one has to check self-consistently that $\psi L_{\rm max}^y f_1(0) \ll f_\infty (0)$.
While this seems fully justified in $d=4$ and $d=5$, it might be slightly less well grounded
in $d=6$. 
Hence, the critical disorder $W_c$ and the absolute value of the exponent $y$ 
might be overestimated in six dimensions.}
We also applied this method in $3d$ (not shown), yielding $W_c = 16.35 \pm 0.1$, $\nu = 1.57 \pm 0.02$, and $y = -1 \pm 0.1$,
in excellent agreement with the results of Refs.~\cite{best3d1,best3d2}. 
Remarkably, the leading irrelevant exponent $y$ seems to depend very 
weakly 
on the spatial dimension 
at least up to $6d$. 

It is remarkable that finite size corrections are governed by scaling variables ($(W-W_c)L^{1/\nu}$ and 
$L^y$) in which the {\it linear} size $L$ enters raised to exponents ($\nu$ and $y$) that seem to have a finite limit when $d\rightarrow \infty$. This suggests a very different behavior from conventional phase transition where scaling variables instead are naturally expressed in terms of $N=L^d$. We will come back to this point 
in the conclusion. 


\subsection{Statistics of level spacings and of wave-functions coefficients} \label{sec:statistics}
 
\begin{figure}
 \includegraphics[angle=0,width=0.48\textwidth]{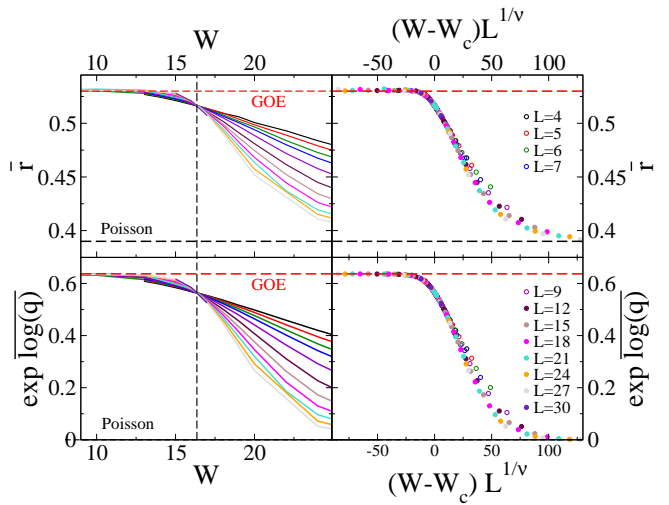}
 \caption{\label{fig:r-q-3d}
 $\overline{r}$ (top-left) and $q^{typ}$ (bottom-left) as a function of the disorder $W$ for several
 system sizes $L$ from $4$ to $30$.
The horizontal dashed lines correspond to the reference GOE and Poisson asymptotic values. The vertical
dashed line spots the position of the AL transition, $W_c \simeq 16.35$. Finite size scaling of the
same data (top and bottom-right panels) showing data collapse obtained for $\nu \simeq 1.57$. Finite-size corrections 
to Eq.~(\ref{eq:one-parameter-scaling}) are observed at small sizes (open symbols), and can be
described by Eq.~(\ref{eq:tps}) with $y \simeq -1$.}
 \end{figure}

In order to analyze the statistics of energy gaps and of wave-functions amplitudes 
we have diagonalized the Hamiltonian~(\ref{eq:H}) for dimensions from $3$ to $6$, 
for several system sizes $L$ (with periodic boundary conditions), 
and for several values of the disorder strength $W$.
For each $L$ and $W$, we have averaged over several realizations of 
the on-site quenched disorder.
Since we are interested in $E=0$, we only focused on $1/16$ of the eigenstates centered around
the middle of the band (we have checked that taking $1/32$ or $1/64$ of the states does
not affect the results, but yields a poorer statistics).
The computer time required for ED grows as the third power of the total number 
of sites of the matrix, $L^{3d}$. 
As a consequence, we can access slightly smaller system sizes with respect to the TM method.
Still, one can simulate rather large values of $L$ for low enough dimensions 
(e.g., $L_{\textrm{max}} = 30$ for $d=3$ and
$L_{\textrm{max}} = 13$ for $d=4$), whereas one is instead limited to very small sizes as
dimensionality is increased ($L_{\textrm{max}} = 8$ for $d=5$ and
$L_{\textrm{max}} = 5$ for $d=6$). 
Note, however, that ED algorithms are faster if one only 
computes the eigenvalues and {\it not} eigenvectors. 
For this reason, in $d=6$ we have been able to obtain some data for the statistics of energy gaps,
for which the knowledge of the eigenfunctions is not necessary, 
also for $L_{\rm max} =6$. 

\begin{figure}
 \includegraphics[angle=0,width=0.48\textwidth]{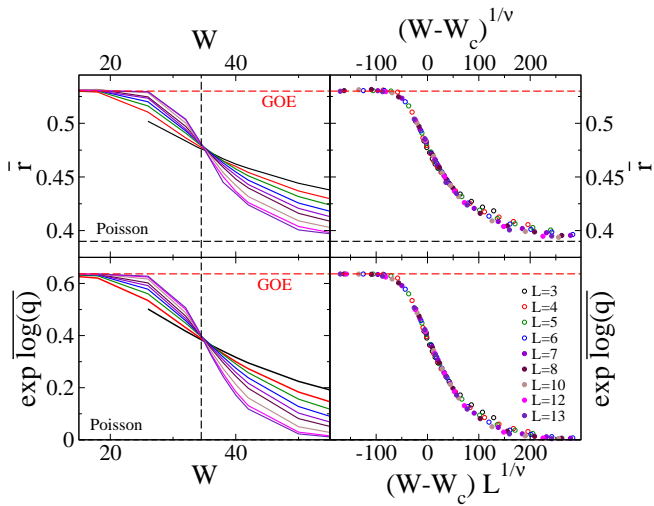}
 \caption{\label{fig:r-q-4d}
 $\overline{r}$ (top-left) and $q^{typ}$ (bottom-left) as a function of the disorder $W$ for several
 system sizes $L$ from $3$ to $13$.
The horizontal dashed lines correspond to the reference GOE and Poisson asymptotic values. The vertical
dashed line spots the position of the AL transition, $W_c \simeq 34.5$. Finite size scaling of the
same data (top and bottom-right panels) showing data collapse obtained for $\nu \simeq 1.11$.
Finite-size corrections to Eq.~(\ref{eq:one-parameter-scaling}) are
observed at small sizes (open symbols), and can be
described by Eq.~(\ref{eq:tps}) with $y \simeq -1$.}
 \end{figure}

We have studied the statistics of
level spacings of neighboring eigenvalues: $s_n = E_{n+1} - E_n \ge 0$,
where $E_n$ is the energy of the $n$-th eigenstate in the sample.
In the extended regime level crossings are forbidden. Hence the eigenvalues are strongly correlated and
the level statistics is expected to be described by RMT 
(more precisely, several results support a general
relationship between delocalization and the Wigner's surmise of the GOE).
Conversely, in the localized phase
wave-functions close in energy are exponentially localized on very distant sites and do not overlap. Thus
there is no level-repulsion and eigenvalues should be distributed similarly to random points thrown on a line
(Poisson statistics).
In order to avoid difficulties
related to the unfolding of the spectrum, we follow~\cite{huse} and measure the ratio of adjacent gaps,
\begin{displaymath}
r_n = \frac{\min \{ s_n, s_{n+1} \}}{\max \{ s_n, s_{n+1} \}} \, ,
\end{displaymath}
and obtain the
probability distribution $\Pi (r)$, which displays a universal form depending on
the level statistics~\cite{huse}. In particular $\Pi (r)$ is expected to converge to its GOE and
Poisson counterpart in the extended and localized regime~\cite{huse,Pr-GOE}, allowing to discriminate between the two phases
as $\overline{r}$ changes from $\overline{r}_{GOE} \simeq 0.5307$ to $\overline{r}_P \simeq 0.3863$
respectively.

\begin{figure}
 \includegraphics[angle=0,width=0.48\textwidth]{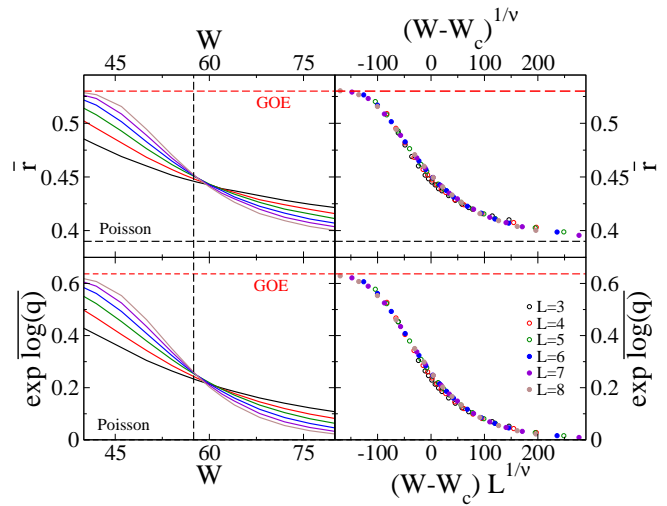}
 \caption{\label{fig:r-q-5d}
 $\overline{r}$ (top-left) and $q^{typ}$ (bottom-left) as a function of the disorder $W$ for several
 system sizes $L$ from $3$ to $8$.
The horizontal dashed lines correspond to the reference GOE and Poisson asymptotic values. The vertical
dashed line spots the position of the AL transition, $W_c \simeq 57.5$. Finite size scaling of the
same data (top and bottom-right panels) showing data collapse obtained for $\nu \simeq 0.96 $.
Finite-size corrections 
to Eq.~(\ref{eq:one-parameter-scaling}) are observed at small sizes (open symbols), and can be
described by Eq.~(\ref{eq:tps}) with $y \simeq -1.2$.}
 \end{figure}

The GOE-Poisson transition can also be captured by correlations between nearby eigenstates such as
the mutual overlap between two subsequent eigenvectors, defined as:
\begin{displaymath} 
q_n = \sum_{i=1}^N | \braket{i}{n} | | \braket{i}{n+1} | \, .
\end{displaymath}
In the GOE regime the wave-functions amplitudes are i.i.d.~Gaussian random variables
of zero mean and variance $1/N$~\cite{porter-thomas}, hence $\overline{q}$
converges to $\overline{q}_{GOE} = 2/\pi$.
Conversely in the localized phase two successive eigenvector are generically peaked around very distant sites and do not overlap, and therefore
$\overline{q}_{P} \to 0$ for $L \to \infty$.
At first sight this quantity seems to be related to the statistics of wave-functions coefficients rather than to energy gaps.
Nonetheless, in all the random matrix models that have been considered in the literature up to now, 
one empirically finds that $\overline{q}$ is directly associated to the statistics of 
level spacings.
The best example of that is provided by the generalization of the Rosenzweig-Porter
random  matrix  model of~\cite{kravtsov}, where there is a whole region of the parameter space 
where wave-functions are
delocalized but multifractal and strongly correlated, while the statistics of neighboring gaps is still described by the GOE ensemble.
In this case one numerically finds that $\overline{q}$ converges to its GOE
universal value $2/\pi$ irrespective of the fact that wave-functions amplitudes are {\it not} i.i.d. Gaussian random variables of variance $1/N$.
 
\begin{figure}
 \includegraphics[angle=0,width=0.48\textwidth]{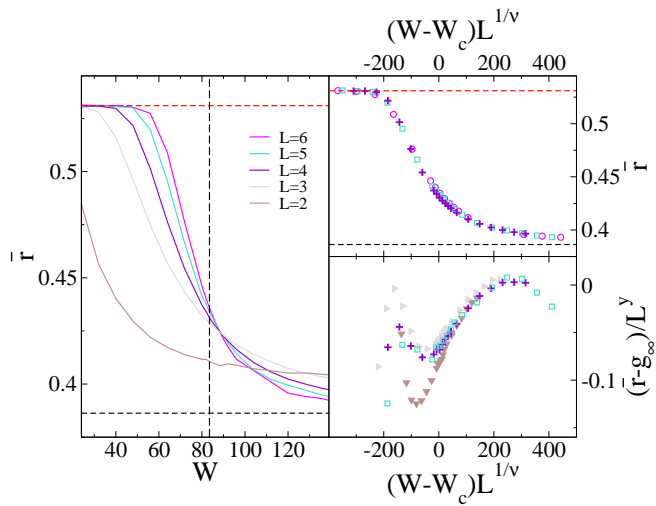}
 \caption{\label{fig:r-q-6d}
 Left panel: $\overline{r}$ as a function of the disorder $W$ for several
 system sizes $L$ from $2$ to $6$. 
The horizontal dashed lines correspond to the reference GOE and Poisson asymptotic values. The vertical
dashed line spots the position of the AL transition, $W_c \simeq 83.5$.
Top-right panel: Finite size scaling of the same data for the largest system sizes only, $L=4$, $5$, and $6$,
showing data collapse for $\nu \simeq 0.84$. Bottom-right panel: 
$\psi g_1 = (\overline{r} - g_\infty)/L^y$ as a function of the
scaling variable $(W-W_c) L^{1/\nu}$ for different sizes $L$ from $2$ to $5$,
showing a reasonably good data collapse for the same value as before of $W_c$ and $\nu$,
and for $y \simeq -1.4$.}
 \end{figure}

In figs.~\ref{fig:r-q-3d}, \ref{fig:r-q-4d}, and \ref{fig:r-q-5d} 
we show the behavior of the average value of the ratio of adjacent gaps, 
$\overline{r}$, and of 
the typical value of the mutual overlap between subsequent eigenvectors,
 $q^{typ} = \exp[\overline{\log q}]$,
 as a function of the disorder $W$, for several system sizes $L$, and for $d=3$, $4$, and $5$ 
respectively. 
As expected, for small (resp. large) enough disorder we recover the universal values
$\overline{r}_{GOE} \simeq 0.5307$ and $q_{GOE}^{typ} = 2 / \pi$ (resp. $\overline{r}_{P} \simeq 0.3863$
and $q_P^{typ} \to 0$) corresponding to GOE (resp. Poisson) statistics.
Data for different system sizes exhibit a crossing point around the critical points $W_c$, which  
coincide, within our numerical accuracy, 
with the ones obtained in the previous subsection from the analysis of the Lyapunov exponent,
and are in good agreement with the ones reported in the literature~\cite{best3d1,best3d2,numerics4d5d}. 
One also finds that for large enough $L$ the whole probability distribution $\Pi (r)$ converges to its GOE and Poisson 
counterparts for $W < W_c$ and $W > W_c$ respectively.
In the right panels of figs.~\ref{fig:r-q-3d}, \ref{fig:r-q-4d}, and \ref{fig:r-q-5d}, we
show that for the largest accessible system sizes the dependence of $\overline{r}$ and $q^{typ}$
on $W$ and $L$ can be described in terms of the scaling functions:
\begin{equation} \label{eq:one-parameter-scaling}
\begin{split}
& \overline{r} (W,L) = g_\infty \! \left( \! w L^{1/\nu} \right) \, , \\
& q^{typ} (W,L) = h_\infty \! \left( \! w L^{1/\nu} \right) \, ,
\end{split} 
\end{equation} 
with $w=(W-W_c)/W_c$. 
The values of $\nu$ are consistent, within our numerical incertitude, with the ones estimated using the TM method 
in the previous subsection, and are in perfect agreement with Refs.~\cite{best3d1,best3d2,numerics4d5d}. 
Deviations from Eq.~(\ref{eq:one-parameter-scaling}) due to FSE 
are clearly visible at small $L$, and can be described 
in terms of systematic corrections to the one-parameter scaling due to  
the presence of irrelevant scaling variables as explained above [see Eq.~(\ref{eq:tps})].
The numerical values of the exponent $y$ describing finite-size corrections to scaling for $\overline{r}$ and $q^{typ}$ 
are compatible, within our numerical precision, with the ones reported in Eq.~(\ref{eq:critical}), 
confirming that the same sets of critical parameters describe
the critical properties of level statistics and transport properties.

As already pointed out before, FSE get stronger as dimensionality is increased. This effect is even more visible when level
statistics is considered. 
In the left panel of fig.~\ref{fig:r-q-6d} we show the behavior of $\overline{r}$ as a function of the disorder strength $W$,
for $L$ from $2$ to $6$ in six dimensions, showing dramatic FSE: The crossing point shifts towards smaller values of $W$ 
from about $W \sim 130$ to $W \sim 86$ as $L$ is increased from $2$ to $6$, and it has not converged yet to $W_c$ even for
the largest available system size. Nevertheless, taking care carefully of finite-size corrections as 
explained in Sec.~\ref{sec:lyapunov}, one is able
to show that the same set of critical parameters found from the analysis of the Lyapunov exponent ($W_c \simeq 83.5$, $\nu \simeq 0.84$, 
and $y \simeq -1.5$) yield a reasonably good finite-size scaling. This is demonstrated by the top-right and bottom-right panels
of fig.~\ref{fig:r-q-6d}, where the scaling functions $g_\infty$ and ($\psi$ times) $g_1$ are found from the data collapse of
the numerical data in terms of the scaling variables $(W - W_c) L^{1/\nu}$. 
(We were not able to repeat the same analysis for $q^{typ}$, since numerical data for the overlap between subsequent eigenvectors
are available only up to $L = 5$.)

\begin{figure}
 \includegraphics[angle=0,width=0.44\textwidth]{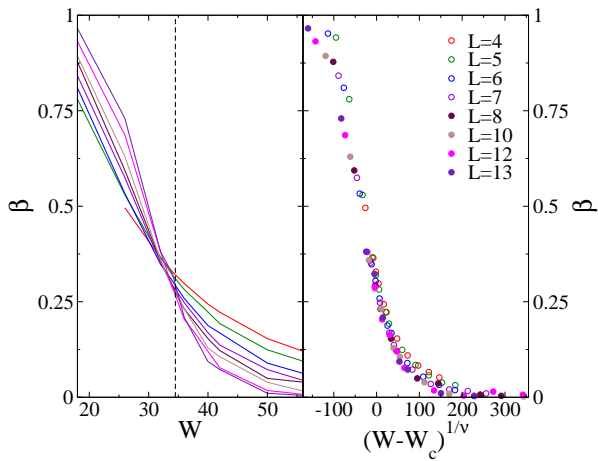}
 \caption{\label{fig:IPR-4d} Left panel: Flowing fractal exponent $\beta$ describing the scaling of the typical value of
the IPR with the system size. The vertical dashed black line corresponds to the critical disorder $W_c \simeq 34.5$.
Right panel: Finite size scaling of the same data showing a reasonably good data collapse
obtained for $\nu \simeq 1.11$. Strong finite-size corrections to the one-parameter scaling are observed at small
sizes (open symbols), and can be
described by Eq.~(\ref{eq:tps}) with $y \simeq -1$.}
 \end{figure}

Analyzing fluctuations of eigenfunctions, we also focused on the (averaged) Inverse Participation Ratio.
The IPR of the eigenfunction $\ket{n}$ is defined
as $\Upsilon_{2}^{(n)} = \sum_{i=1}^{L^d} | \braket{i}{n} |^4$.
In the full extended regime wave-functions are uniformly spread over all the volume,
thus $\braket{i}{n}$ are random variables of order $1/\sqrt{L^d}$, due to normalization, and
$\overline{\Upsilon_2}$ vanishes as $C/L^d$ for $L \to \infty$---the prefactor $C$ depends on the
disorder strength $W$, approaching its GOE value equal to $3$ deep in the metallic phase.
Conversely in the localized phase wave-functions are localized on $O(\xi^d)$ sites and
$\overline{\Upsilon_2}$ approaches a constant value in the thermodynamic limit
(in the infinite disorder limit, $W \to \infty$, one has that $\overline{\Upsilon_2} \to 1$).


From the wave-functions amplitudes obtained via ED, we have computed
the typical value of the IPR, defined as $\Upsilon_2^{typ} = \exp [ \overline{ \log \Upsilon_2} ]$,
for several values of the disorder strength and of the system size $L$, and for dimensions from $3$ to $5$.
The flowing fractal exponent 
$\beta$ 
describing the scaling of $\Upsilon_2^{typ}$ 
with $L$ can then be approximately evaluated as:
\begin{equation} \label{eq:exp-IPR-SS}
\begin{split}
\beta (W,L) & = - \, \frac{\log \Upsilon_2^{typ} (W,L) - \log \Upsilon_2^{typ} (W,L-1)}{d[\log L - \log (L-1)]} \, . 
\end{split}
\end{equation}
In fig.~\ref{fig:IPR-4d} we plot the numerical results for the 
exponent $\beta$ as a function of $W$ for several system sizes
in four dimensions, showing a similar---although much less clean---behavior compared to the one
found for the statistics of energy gaps:
For $W<W_c$ one observes that $\beta$ grows with $L$; its behavior is compatible 
with an approach towards $1$ for $L$ large enough, corresponding to full delocalized wave-functions.
Conversely, for $W > W_c$ the exponent $\beta$ 
decreases as the system size is increased, and seems to approach $0$ for large $L$,  
implying that $\Upsilon_{2}^{typ} \to \textrm{cst}$, as expected for localized eigenstates.
For the largest available sizes, the curves corresponding to different values of $L$ cross approximately around $W_c \simeq 34.5$.
Although $\beta$ is affected by much larger fluctuations and stronger FSE compared to $\overline{r}$ and $q^{typ}$, 
the same set of critical parameters found before ($W_c \simeq 34.5$, $\nu \simeq 1.11$, and $y \simeq -1$) 
yields a reasonably good data collapse of numerical data, 
as shown in the right panel of fig.~\ref{fig:IPR-4d}.
Similar results are also found in dimensions $3$ and $5$ (not shown).
This analysis can not be performed in six dimensions, due to the fact that the IPR can be 
measured only up to $L_{\rm max} = 5$, which is not sufficiently large to take care in an accurate way
of the strong FSE.

\begin{figure}
 \includegraphics[angle=0,width=0.46\textwidth]{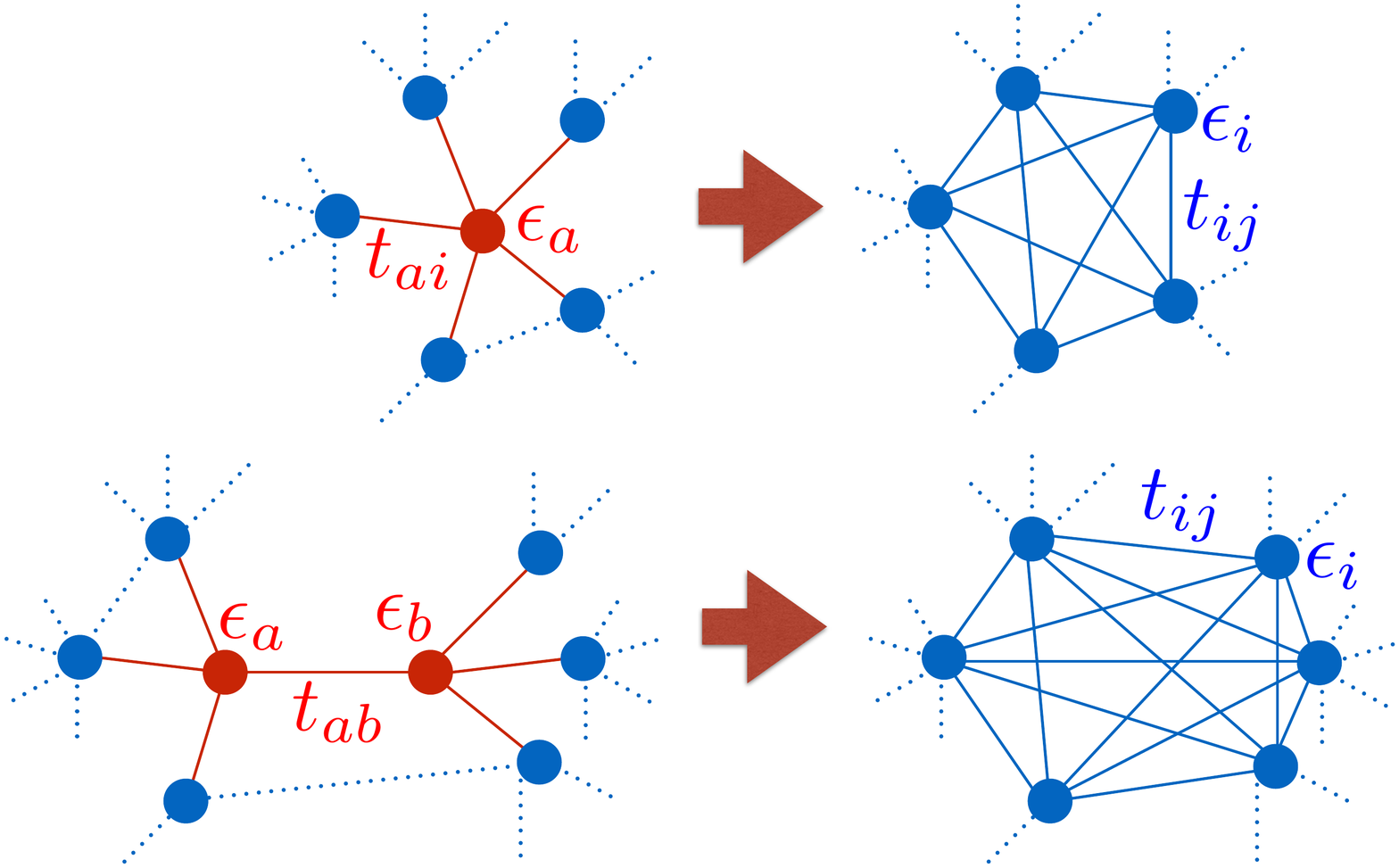}
 \caption{\label{fig:sdrg-sketch}
Sketch of the SDRG decimation procedure for a site (top), and a bond (bottom) transformation.
Dotted blue lines represent pre-existing hopping amplitudes before decimation. Solid blue lines represent
new or renormalized bonds. The on site energies of all the neighbors of the decimated sites (blue circles)
are renormalized as well.} 
 \end{figure}

\section{Strong Disorder RG} \label{sec:SDRG}

In this section we present our results based on the Strong Disorder RG approach
for AL recently introduced in~\cite{sdrg,sdrg1d}.
The SDRG is an efficient real-space decimation procedure, consisting in integrating-out iteratively the
largest coupling constant in the Hamiltonian.
The ideas behind this method reside in the seminal work of Ref.~\cite{sdrg-fisher}, 
and have been successful applied to
describe the critical and near-critical behavior of the
Random Transverse-Field Ising model and other random
magnetic transitions~\cite{sdrg-RFIM}, and have also been recently used
in electronic systems~\cite{sdrg-electrons}.

In the case in which the strongest energy scale happens to be the on-site energy $|\epsilon_a|$ on site $a$, 
as sketched in the top panel of fig.~\ref{fig:sdrg-sketch}, one can
perform the Gaussian integral over $\phi_a$ in Eq.~(\ref{eq:Z}),
obtaining 
a RG transformation for the on-site energies on all the neighbors $i$
of $a$ and for the hopping amplitudes between all possible pairs of neighbors $(ij)$ of $a$:
\begin{equation} \label{eq:SDRG_e}
\begin{split}
\epsilon_i & \to \epsilon_i - \frac{t_{ai}^2}{\epsilon_a} \, , \\
t_{ij} & \to t_{ij} - \frac{t_{ai} t_{aj}}{\epsilon_a} \, .
\end{split}
\end{equation}
Similarly, if the strongest energy scale is the hopping amplitude $|t_{ab}|$ between 
sites $a$ and $b$, as sketched in the bottom panel of fig.~\ref{fig:sdrg-sketch}, 
performing the Gaussian integrals over $\phi_a$ and $\phi_b$
in Eq.~(\ref{eq:Z})
yields the following 
RG transformation for the on-site energies on all the neighbors $i$
of $a$ and $b$ and for the hopping amplitudes between all possible pairs of neighbors $(ij)$ of $a$ and/or $b$:
\begin{equation} \label{eq:SDRG_t}
\begin{split}
\epsilon_i & \to \epsilon_i - \frac{\epsilon_a t_{bi}^2 - 2 t_{ab} t_{ai} t_{bi} + \epsilon_b t_{ai}^2}{\epsilon_a \epsilon_b - t_{ab}^2} \, , \\
t_{ij} & \to t_{ij} -  \frac{\epsilon_a t_{bi} t_{bj} - t_{ab} ( t_{ai} t_{bj} + t_{aj} t_{bi} ) + \epsilon_b t_{ai} t_{aj}}{\epsilon_a \epsilon_b - t_{ab}^2}  \, .
\end{split}
\end{equation}
Note that Eqs.~(\ref{eq:SDRG_t}) can be obtained using Eqs.~(\ref{eq:SDRG_e}) twice to eliminate first site $a$ and then site $b$.

Eqs.~(\ref{eq:SDRG_e}) and (\ref{eq:SDRG_t}) 
are in fact exact RG transformations, as it was first shown in~\cite{aoki}. 
However, 
the number of non-zero matrix elements grows very rapidly
under RG due to the proliferation of new bonds (except, of course, in $1d$~\cite{sdrg1d}). 
This makes the numerical analysis unpractical.
Several procedures have been proposed to solve this problem, which is also encountered 
in similar SDRG schemes for electronic systems~\cite{sdrg-electrons}
as well as for other disordered models such as random transverse-field Ising model~\cite{sdrg-RFIM}.
In this work we follow~\cite{sdrg} and set a maximum coordination number $k_{\rm max}$ per site, throwing away most of the weak
couplings. The rationale behind this procedure is that---at least in high enough dimension---the critical properties of AL 
are controlled by a strong disorder limit, and the weak coupling constants generated under RG are in fact ``irrelevant''.

In order to check whether or not this 
assumption is correct, it is important to analyze 
the accuracy of the results obtained using the SDRG and study their convergence with $k_{\rm max}$.
We first focus on the average DOS, 
$\rho = - \mbox{Tr} \, \mbox{Im} {\cal G} / (\pi L^d)$.\\
We define the following quadratic form $\Gamma[\phi_i; \{\omega_i,\sigma_{ij},\kappa\}]$ of the
auxiliary fields $\phi_i$:
\begin{equation} \label{eq:SDRGgamma}
\Gamma[\phi_i; \{\omega_i,\sigma_{ij},\kappa\}] = 
\sum_i \omega_i \phi_i^2 + \sum_{i<j} \sigma_{ij} \phi_i \phi_j + i \kappa \, ,
\end{equation}
in terms of which the average DOS can be written as:
\begin{equation} \label{eq:avDOS}
\begin{split}
\rho &= 
\frac{i}{\pi L^d Z} \, \mbox{Im} \int \prod_{i=1}^N {\rm d} \phi_i \, \\
& \qquad \qquad \times 
\Gamma[\phi_i;\{\omega_i=1,\sigma_{ij}=0,\kappa=0\}]
\, e^{ S[\phi_i] } \, ,
\end{split}
\end{equation}
where $Z$ is defined in Eq.~(\ref{eq:Z}). 
When a site or a bond are integrated-out under the RG transformations, 
some of the coefficients of $\Gamma$ (i.e., those involving the neighboring sites of the decimated variables)
must then be renormalized as well.
Hence, although at the level of the initial conditions one has that 
$\omega_i = 1$ for all $i$, $\sigma_{ij} = 0$ for all $(ij)$, and $\kappa = 0$ [see Eq.~(\ref{eq:avDOS})],
in order to compute the average DOS 
one needs to keep track of the flow of all the 
coefficients of $\Gamma$ 
under RG.
For example, when a given site, say site $a$, is decimated out, one has to renormalize the 
coefficients $\omega_i$ of all sites $i$ neighbors of $a$, the coefficients $\sigma_{ij}$ of 
all possible pairs of neighbors $(ij)$ of $a$, as well as the value of the constant $\kappa$.
This can be easily done by Gaussian integration: 
\begin{equation} \label{eq:SDRGcoeff}
\begin{split}
\omega_i & \to \omega_i + \frac{\omega_a t_{ai}^2}{\epsilon_a^2} - \frac{t_{ai} \sigma_{ai}}{\epsilon_a} \, ,\\
\sigma_{ij} & \to \sigma_{ij} + \frac{2 \omega_a t_{ai} t_{aj}}{\epsilon_a^2} - \frac{t_{ai} \sigma_{aj} + t_{aj} \sigma_{ai}}{\epsilon_a} \, ,\\
\kappa & \to \kappa + \frac{\omega_a}{\epsilon_a} \, .
\end{split}
\end{equation}
Similarly, when the hopping amplitude between sites $a$ and $b$ is eliminated, one can determine 
analogous RG relations for the coefficients of Eq.~(\ref{eq:SDRGgamma}) using 
Eq.~(\ref{eq:SDRGcoeff}) twice, first on site $a$ and then on
site $b$.
At the end of the RG, when all sites have been integrated-out,
$\rho$ can be then obtained from Eq.~(\ref{eq:avDOS}) as (minus) the imaginary part of the final value of $\kappa$ divided by $\pi L^d$.

\begin{figure}
 \includegraphics[angle=0,width=0.44\textwidth]{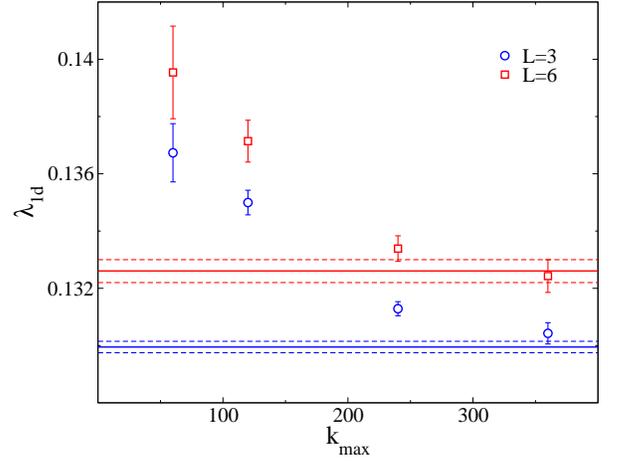}
 \caption{\label{fig:SDRG-lambda1d}
Quasi-$1d$ dimensionless localization length, $\lambda_{1d}$, obtained using the SDRG procedure for different
values of $k_{\rm max}$, at the AL critical point, $W_c \simeq 83.5$, in $6$ dimensions, and
for $L=3$ (blue circles) and $L=6$ (red squares).
The horizontal blue (resp., red) solid and dashed lines corresponds to the average value of $\lambda_{1d}$ and its fluctuations computed
using the TM method for $L=3$ (resp., $L=6$), showing that for 
$k_{\rm max} \gtrsim 240$ the approximate SDRG results converge, within our numerical accuracy, to the
exact values.}
 \end{figure}

We have computed the average DOS around the AL critical points for dimensions from $3$ to $6$ using this method  
for several values of $k_{\rm max}$, and compared its numerical value with the one obtained
from ED, finding an excellent agreement even at small values of $k_{\rm max}$.
In practice, 
already for $k_{\rm max} \gtrsim 60$ the average DOS
obtained via the SDRG coincides within error-bars and sample-by-sample 
with the one computed from ED for all the accessible system sizes and in all dimensions.

We turn now to transport properties. In particular, in the following we compare the results for the
dimensionless quasi-$1d$ localization length computed from  
the TM
approach as described in Sec.~\ref{sec:lyapunov}, with the ones obtained using
the SDRG with
different values of $k_{\rm max}$. 
More precisely, 
we consider the quasi-$1d$ bar of fig.~\ref{fig:quasi-1d} and
instead of solving Eq.~(\ref{eq:cavity-quasi1d}) exactly via LU decomposition, 
we apply the SDRG to invert the matrix $[{\cal G} (x)]^{-1}$ in an approximate way, as 
explained in the following:
\begin{itemize}
\item[(1)] We 
start from the layer at $x=0$ which is in contact with an electron bath ($\eta > 0$) and 
integrate-out progressively all the sites of the layer 
using Eqs.~(\ref{eq:SDRG_e}) and (\ref{eq:SDRG_t}), eliminating iteratively the
strongest energy scale, until no sites are left on the layer. 
\item[(2)] In such a way, at the end of step (1) 
we end up with an approximate expression for the matrix element of the
inverse cavity Green's function on the layer $x=1$, $[{\cal G}(x=1)]^{-1}_{ij}$ 
(in absence of the layer at $x=2$).
Then, knowing the l.h.s. of 
Eq.~(\ref{eq:cavity-quasi1d}), one can infer the matrix elements of the (cavity) Green's function 
on the layer $x=0$, ${\cal G}_{ij} (x=0)$, and compute the typical value of the imaginary part of 
its diagonal elements, $\overline{\log \mbox{Im} {\cal G} (x=0)}$.
\item[(3)] We then integrate-out progressively all the sites of the layer $x=1$ 
using Eqs.~(\ref{eq:SDRG_e}) and (\ref{eq:SDRG_t}) to eliminate iteratively the
strongest energy scale, yielding the matrix elements of the 
inverse cavity Green's function $[{\cal G}(x=2)]^{-1}_{ij}$ on the
layer $x=2$ in absence of the subsequent layer ($x=3$), and use Eq.~(\ref{eq:cavity-quasi1d}) ``backwards''
to infer  ${\cal G}_{ij} (x=1)$. We measure $\overline{\log \mbox{Im} {\cal G} (x=1)}$ and 
repeat the whole process until the layer $x = L_x$ is reached.
\end{itemize}
This procedure allows to compute the dimensionless quasi-$1d$ 
localization length in a considerably faster way compared to exact LU decomposition.
In fig.~\ref{fig:SDRG-lambda1d} we plot the results for $\lambda_{1d}$ at the AL critical point in dimension $6$ 
($W_c \simeq 83.5$) for different values of $k_{\rm max}$ and for $L=3$ and $6$, showing that for $k_{\rm max} \gtrsim 240$
the numerical values of $\lambda_{1d}$ obtained via the SDRG approach converge, within our numerical precision, with the ones
obtained from exact techniques.
Similar results are found in all dimensions down to $d=3$ (not shown). 

This analysis shows that the results obtained using the SDRG approach for both for the average DOS and the Lyapunov exponent 
converge already for reasonably small values of $k_{\rm max}$ 
to the exact ones in all spatial dimensions, 
at least close enough to the AL critical point.\footnote{It is natural to expect that the accuracy of the SDRG gets worse at small
disorder strength, deep into the metallic phase.} 
Hence, 
the critical parameters found using the SDRG approach (for sufficiently large $k_{\rm max}$) 
coincide, within error-bars, with the ones
given in~Eq.~(\ref{eq:critical}).
Since the computer time required for an efficient algorithmic implementation of the SDRG procedure scales as 
$d L^d (\log L) k_{\rm max}^2 (\log k_{\rm max})$, one can in principle apply this method to obtain very accurate results 
for much larger system sizes compared with the exact numerical techniques. 
The SDRG can then also be applied to study AL in dimensions larger than $6$. 
This analysis goes beyond the scope of this work. Preliminary results in this direction 
have already been obtained in~\cite{sdrg} up to $d=10$. 

In the last part of this section, we focus instead on 
the properties of the flow of the SDRG close to the AL critical point. 
More precisely, we study the evolution under RG of the probability distributions of the diagonal and off-diagonal matrix elements,
$Q_\tau (\epsilon)$ and $R_\tau (t)$ respectively---the index $\tau$ corresponds to the RG ``time''.
It is important to stress that these probability distributions 
do not contain {\it all} the relevant physical information on the system.
For instance, they are insensitive to correlations between on-site energies and hopping amplitudes and/or
spatial correlations between matrix elements which may be possibly generated during the flow.
However, as we will discuss below, they can be still used to gather some useful qualitative insights on the critical
properties of AL in high dimension. 

In the following, for simplicity, we will restrict ourselves to the case of real matrix elements (i.e., we set $\eta=0$ 
on all the sites of the system). Similar results are obtained if one considers a finite (but small, e.g. 
$\eta \sim 10/L^d$) imaginary
regulator and study, for instance, the flow of the probability distributions of the modulus of diagonal and off-diagonal
matrix elements.
At the AL critical point, 
the initial conditions for the probability distributions of on-site energies and hopping amplitudes are:
\begin{equation} \label{eq:SDRGic}
\begin{split}
Q_{\tau = 0} (\epsilon) & = \frac{1}{W_c} \, \theta \! \left ( \frac{W_c}{2} - | \epsilon | \right) \, , \\
R_{\tau = 0} (t) & = \frac{2d}{N-1} \, \delta(t-1) + \frac{N - 1 - 2d}{N-1} \, \delta(t) \, .
\end{split}
\end{equation}
The critical disorder $W_c$ is much larger than $1$ already in three
dimensions---and it grows very fast as $d$ is increased (see fig.~\ref{fig:Wc}).
As a consequence, at the beginning of the RG, the strongest energy scales are provided by the 
sites with on-site energies close to the edges of the support of $Q_{\tau = 0} (\epsilon)$.
As these sites are integrated-out, new hopping amplitudes are generated, and the two $\delta$-peaks of
$R_{\tau = 0} (t)$ acquire a finite support.
Hence, as the RG time $\tau$ grows, $Q_{\tau} (\epsilon)$ shrinks and $R_{\tau} (t)$ broadens.
When the support of the two distributions become approximately the same, 
we observe a stationary state.\footnote{In practice, we observe that in all dimensions from $3$ to $6$ 
this happens
when the support of the probability distributions of the diagonal and off-diagonal elements become 
of $O(1)$. As $\tau$ is further increased, the number of matrix elements left in the systems becomes very small
and the stationary distribution is wiped out. However, this is a finite-size effect
which could in principle be avoided taking larger and larger systems.}

\begin{figure}
 \includegraphics[angle=0,width=0.48\textwidth]{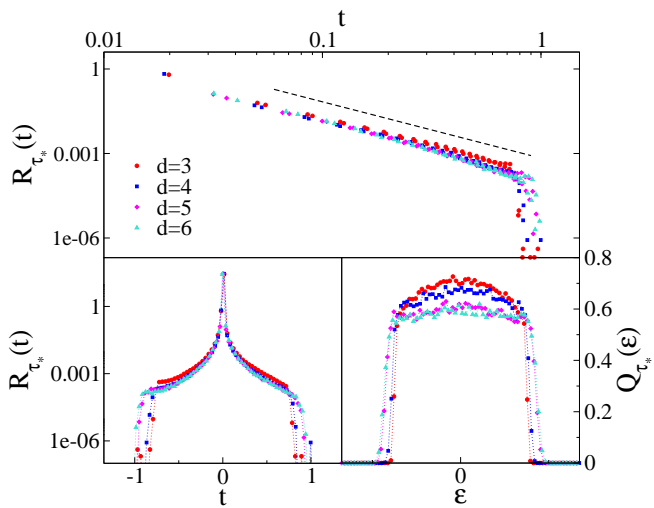}
 \caption{\label{fig:RGflow}
Bottom-left and bottom-right panels: The stationary distributions $R_{\tau_\star} (t)$ and $Q_{\tau_\star} (\epsilon)$  
at the AL critical points
in dimensions from $3$ to $6$. The system size is $L=33$ in $3d$, $L=14$ in $4d$, $L=8$ in $5d$, and $L=6$ in $6d$, in such a way
that the total number of sites is approximately the same, $N \sim 4 \cdot 10^4$, in all dimensions.
The stationary state is reached for a RG time $\tau_\star$ such that the number of sites left in the system are approximately $1/8$ of the
initial ones in $3d$, $1/16$ in $4d$, $1/26$ in $5d$, and $1/40$ in $6d$. The value of $k_{\rm max}$ is set to $360$ in all dimensions.
Top-panel: The same data of the bottom-left panel plotted in a log-log scale, showing the power law behavior of 
$R_{\tau_\star} (t) ~ \sim t^{- \gamma}$ with $\gamma \simeq 2$ (black dashed line), for $t$ smaller than a cut-off of $O(1)$.}
 \end{figure}

The stationary distributions $Q_{\tau_\star} (\epsilon)$ and $R_{\tau_\star} (t)$ at the AL critical points
are plotted in fig.~\ref{fig:RGflow}.
Despite the fact that the initial conditions (\ref{eq:SDRGic}) change dramatically as $d$ is increased,
we observe that
$Q_{\tau_\star} (\epsilon)$ and $R_{\tau_\star} (t)$ are strikingly similar in all spatial dimensions
from $3$ to $6$.
This implies that the RG flow, and thus the critical properties of AL, are controlled by a fixed point which 
is very similar for all $d \ge 3$.
As shown in the inset of fig.~\ref{fig:RGflow}, the tails of $R_{\star} (t)$ seems to
be described by a 
power law, $R_{\tau_\star} (t) \sim t^{-\gamma}$, with an exponent $\gamma \simeq 2$ which is
also roughly independent on $d$, and a cut-off for hopping amplitudes of $O(1)$, which seems
to drift slowly to larger values of $t$ as $d$ is increased. 
Note however that for large $d$ the initial conditions~(\ref{eq:SDRGic}) get further and further from
the stationary distributions. One needs then more and more RG steps to approach the stationary
regime of the flow, i.e., $\tau_\star$ increases as $d$ grows. 
For this reason, FSE on $Q_{\tau_\star} (\epsilon)$ and $R_{\tau_\star} (t)$
also increase as $d$ is increased since         
for $\tau = \tau_\star$ we are left with smaller systems and fewer matrix elements (see the caption of
fig.~\ref{fig:RGflow} for more details).

The power law tails of $R_{\tau_\star} (t)$ are reminiscent of a strong disorder 
fixed point scenario~\cite{irfp}, since they are related to the divergence of the variance
of the distribution of the hopping amplitudes.
However, the fact that the stationary distributions exhibit a cut-off on a scale of 
$O(1)$---which does not seems
to be due to a FSE---implies that in fact all matrix elements stay of $O(1)$ and that the
disorder do not grows under iterations of the RG transformations.
Nonetheless, although at any finite $d$ the fixed point is not of a ``truly'' infinite disorder type, 
the SDRG approach still provides an efficient and accurate approximation scheme.
Furthermore, we find that the cut-off on the power law hopping distribution increases and possibly diverges 
in the large-$d$ limit, suggests that in this case one 
recovers a genuine IRFP scenario~\cite{irfp}.

All in all, 
these observations provide a convincing indication of the fact that the properties of AL in high dimensions
are governed by a ``strong disorder'' fixed point, as already suggested in~\cite{mirlin1,sdrg}.
This idea is also supported by the results of the SUSY approach for the critical properties of AL on tree-like structures and
infinite dimensional models~\cite{infinitedexact,SUSY-tree}, and will be discussed 
in more details in the next section.

\begin{figure}
 \includegraphics[angle=0,width=0.44\textwidth]{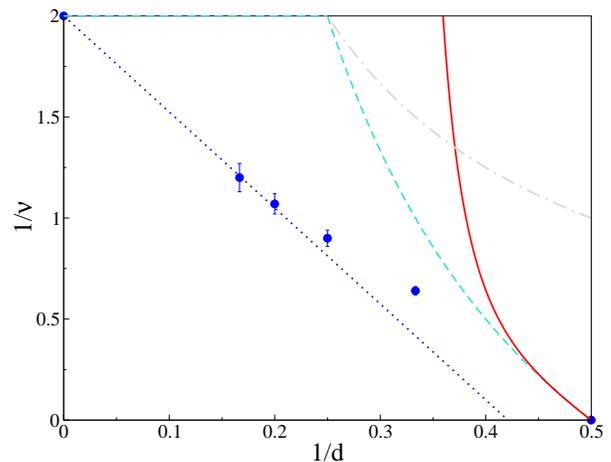}
 \caption{\label{fig:nu}
Numerical values of the inverse of the critical exponent $\nu$ as a function of $1/d$ in dimensions from $3$ to $6$ (blue circles), 
showing a smooth behavior interpolating from $\nu \to \infty$ in $d=2$ to $\nu = 1/2$ in $d \to \infty$~\cite{SUSYdinf}.
The turquoise dashed line shows the predictions of the self-consistent theory of~\cite{vollhardt}, with $d_U=4$.
The dashed-dotted magenta line corresponds to the lower bound $\nu \ge 2/d$ provided by the Harris
criterion~\cite{harris}.
The red solid line shows the dimensional dependence of $\nu$ obtained from a perturbative analysis of the NL$\sigma$M to five-loop 
in $\epsilon = d - 2$~\cite{5-loops}, Eq.~(\ref{eq:5-loops}). The straight dotted blue 
line is a linear fit corresponding to the first correction in $1/d$ from which we find 
$1/\nu\simeq 2- 4,75/d$.}
 \end{figure}

\section{Weak versus strong coupling: analysis of dimensional dependence} \label{sec:results}

In this section we analyze the dimensional dependence at criticality 
of all observables discussed previously. As we shall show, 
approaching the lower critical dimension, $d_L=2$, the critical point corresponds 
to weak disorder (or, equivalently, weak coupling
in terms of 
the NL$\sigma$M): when $d$ approaches two the system at criticality is more and more 
metallic-like and described by the GOE universality class. On the contrary, when $d\rightarrow \infty$,
the system at criticality is more and more
insulating-like and described by the Poisson universality class. This section presents results supporting one of the main message of this work, which is that the infinite dimensional limit is a better starting point to describe systems in all dimensions down to $d=3$. 
\subsubsection{Critical exponents and level statistics} 
We start by focusing on the critical exponent $\nu$, whose behavior as a function of $1/d$ is
plotted in fig.~\ref{fig:nu}.
One clearly observes that $\nu$ continuously decreases from $\nu \to \infty$ in $d=2$ to the value 
$\nu = 1/2$ in $d \to \infty$ predicted by the SUSY approach~\cite{SUSYdinf}, showing no sign of saturation.
This strongly indicates that the upper critical dimension of AL is infinite, as already suggested in~\cite{duinfinite,garcia},
in contrast, for instance,
with the self-consistent theory of~\cite{vollhardt}, which predicts $d_U = 4$ (turquoise dashed line).
The perturbative analysis of the effective field theory based on the replicated NL$\sigma$M has been
carried to five-loops order in $\epsilon = d-2$~\cite{5-loops}, yielding:
\begin{equation} \label{eq:5-loops}
\nu = \frac{1}{\epsilon} - \frac{9}{4} \zeta(3) \epsilon^2 + \frac{27}{16} \zeta(4) \epsilon^3
+ O(\epsilon^4) \, .
\end{equation}
Such dimensional dependence of the critical exponent corresponds to  the solid red line of
fig.~\ref{fig:nu}, and yields a very poor agreement with the numerical results even in low dimensions.
In fact, Eq.~(\ref{eq:5-loops}) violates the lower bound
$\nu \ge 2/d$ based on the Harris criterion~\cite{harris} (dashed-dotted magenta curve)
already in $3d$.
The straight dotted blue 
line in fig.~\ref{fig:nu} is a linear fit corresponding to the first correction in $1/d$ from which we find 
\begin{displaymath}
\frac{1}{\nu} \simeq 2-\frac{4,75}{d} \, .
\end{displaymath} 
The quality of the fit shows that the first correction in $1/d$ performs much better than the expansion to the fifth order in $d-2$ down to $d=3$.     

\begin{figure}
 \includegraphics[angle=0,width=0.48\textwidth]{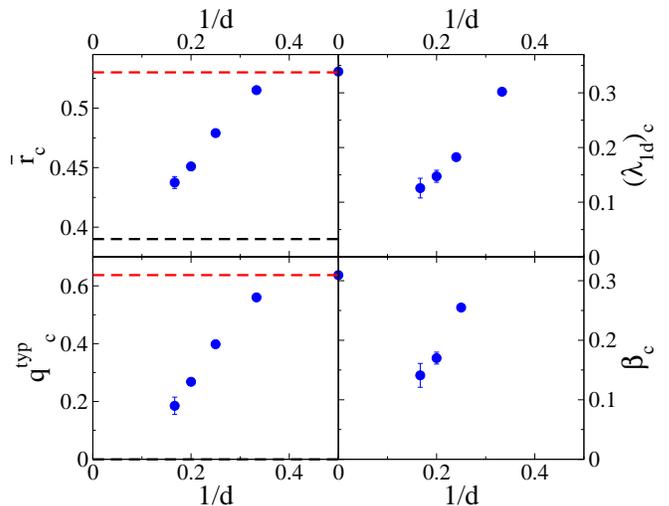}
 \caption{\label{fig:rc}
Dimensional dependence of $\overline{r}_c$ (top-left panel), $q^{typ}_c$ (bottom-left panel), $(\lambda_{1d})_c$ (top-right panel),
and $\beta_c$ at the AL critical point as a function of $1/d$. The dashed horizontal red (resp. black) lines
correspond to the reference GOE (resp. Poisson values).}
 \end{figure}

As mentioned above, these observations suggest that
the critical properties of AL away from the lower critical dimension 
might be governed by a strong disorder regime, as suggested in~\cite{mirlin1,sdrg}.
This idea is fully confirmed by the analysis of the 
critical values and their dimensional dependence:
In fig.~\ref{fig:rc} we plot $\overline{r}_c$ 
(top-left panel), 
$q^{typ}_c$ 
(bottom-left panel), $(\lambda_{1d})_c$ 
(top-right panel), 
and $\beta_c$ 
(bottom-right panel) as a function of $1/d$. 
In $d=2 + \epsilon$ dimensions 
the critical point corresponds to weak disorder (or, equivalently, weak coupling
in terms of 
the NL$\sigma$M), which means that the critical level statistics is close
to the GOE one.
With increasing $d$ the critical point moves continuously towards strong disorder (strong coupling), and
$\overline{r}_c$ and $q^{typ}_c$ approach the Poisson reference values, 
suggesting that the critical level statistics in the in infinite dimensional limit is
of Poisson form, like in the localized phase.
Similarly, $\beta_c$ decreases as $d$ is increased and seems to vanish in the $d \to \infty$ limit, 
implying that the IPR has a finite limit at the AL critical point in infinite dimensions, as predicted by the SUSY approach~\cite{SUSY-tree}.
Finally, $(\lambda_{1d})_c$ is also a decreasing function of $d$, and smoothly approaches $0$ for $d \to \infty$, 
confirming the idea that the AL critical point in infinite dimensions 
is strongly localized also as far as transport properties are concerned.

\subsubsection{Dimensional dependence of the critical disorder strength}
\begin{figure}
 \includegraphics[angle=0,width=0.44\textwidth]{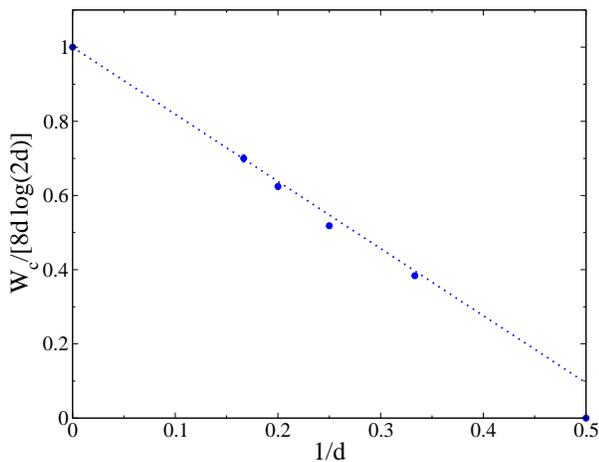}
 \caption{\label{fig:Wc}
Dimensional dependence of the critical value of the disorder strength, $W_c$, divided by $8 d \log (2 d)$,
which corresponds to the exact asymptotic behavior 
on tree-like lattices in the large connectivity limit~\cite{victor}. The straight dotted blue line is a fit that takes into account corrections in $1/d$, from which we find: $W_c/8 d \log (2 d)\simeq 1 - 1,81/d$.}
 \end{figure}

It is also interesting to study the  
dimensional dependence of the critical value of the disorder strength $W_c$.
Fig.~\ref{fig:Wc} 
shows that $W_c$ grows faster than $d$ (which would be the natural scale 
set by the coordination number
for conventional phase transitions) as the dimensionality is increased and 
seems to approach the curve $W_c/t \sim 4(2d-1) \ln(2d - 1)$ for large $d$, which corresponds to the
exact asymptotic behavior on tree-like structures in the large connectivity limit~\cite{abou,victor}.
As for $\nu$, we numerically evaluated corrections to the $d\rightarrow \infty$ result by a fit containing 
the first correction in $1/d$ (straight dotted blue line in fig.~\ref{fig:Wc}). As before, the first correction 
performs impressively well down to $d=3$.  


\subsubsection{Multifractality}
\begin{figure}
 \includegraphics[angle=0,width=0.44\textwidth]{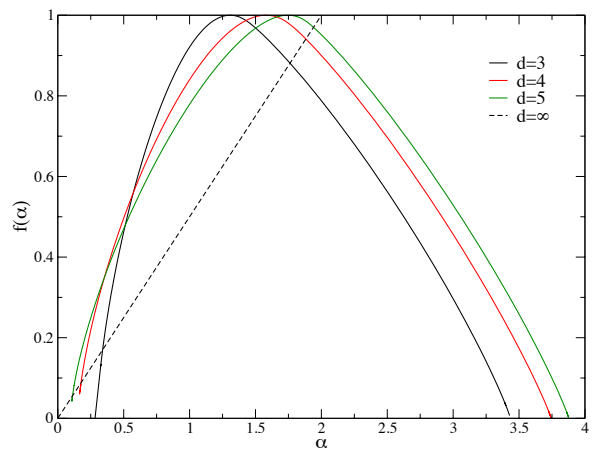}
 \caption{\label{fig:multi}
Re-scaled singularity spectrum $f(\alpha)/d$ as a function of $\alpha/d$ at the AL critical point in
dimensions from $3$ to $5$.
The dashed black straight line $f(\alpha) = \alpha/2$ for $\alpha \in [0,2]$ corresponds to
the prediction of~\cite{evers} in the $d \to \infty$ limit.}
 \end{figure}

We finally turn to the analysis of the dimensional dependence of the critical multifractal spectra of wave-function
amplitudes. Having in our disposition all the coefficients of the eigenvectors from ED, we can easily find
the scaling behavior of all moments
\begin{displaymath}
\Upsilon^{(n)}_q = \sum_{i=1}^{L^d} | \braket{i}{n} |^{2q} \propto L^{-\tau(q)} \, ,
\end{displaymath}
with the system size $L$. (Note that $\overline{\Upsilon_1} = 1$ due to the normalization condition, and
$\overline{\Upsilon_2}$ is the IPR studied in Sec.~\ref{sec:statistics}.)
In the metallic phase the wave-functions amplitudes are of $O(1/L^d)$ and 
$\tau(q) = dq - 1$, whereas $\tau(q) = 0$ in the insulating regime. 
At criticality $\tau(q)$ is characterized by anomalous scaling exponents~\cite{evers}
which are the signatures of multifractal states.
It is customary to introduce the singularity spectrum $f(\alpha)$, which
denotes the fractal dimension of the set of points where the wave-function amplitude is
$| \braket{i}{n} |^2 \sim L^{-\alpha}$ (in our discrete system the number of such points $N (\alpha)$
scales as $L^{f (\alpha)}$):
\begin{displaymath}
\Upsilon_q = \sum_{i=1}^N | \braket{i}{n} |^{2q} \sim \int \textrm{d} \alpha \exp \left[ \left( f(\alpha) 
- q \alpha \right) \log L \right] \, .
\end{displaymath}
Then, in the thermodynamic limit, the saddle point computation of $\Upsilon_q$ leads to the following Legendre transformation:
\begin{equation} \label{eq:legendre}
\begin{split}
& \alpha = \frac{{\rm d} \tau (q)}{{\rm d} q} \, , \qquad q = f^\prime (\alpha) \, , \\
& f(\alpha) = \alpha q - \tau(q) \, .
\end{split}
\end{equation}
$f(\alpha)$ is by definition a convex function of $\alpha$.
The value $q=0$ is associated with the most probable value $\alpha_{m}$
of the wave-function coefficients, where the singularity spectrum
reaches its maximum, $f(\alpha_{m}) = d$. The value $q=1$ is associated
with the point $\alpha_1$ such that $f(\alpha_1) = \alpha_1$, and $f^\prime(\alpha_1) = 1$.
A finite support $0 < \alpha_- < \alpha < \alpha_+$ where $f(\alpha)>0 $ in the $L \to \infty$ limit 
is a signature of multifractality, while for ergodic states,
$f(\alpha) = - \infty$ unless for $\alpha = d$, where $f(d) = d$. 
From the ED data we have evaluated the typical value of the exponent $\tau(q)$ at the AL critical point as~\cite{multi3d}:
\begin{displaymath}
\tau_q^{typ} = - \frac{{\rm d} \, \overline{\log \Upsilon_q}}{{\rm d} L} \, ,
\end{displaymath}
from which the spectrum of fractal dimensions $f(\alpha)$ can be determined 
applying the Legendre transformation, Eq.~(\ref{eq:legendre}).
Our numerical results in dimensions from $3$ to $5$ are plotted in fig.~\ref{fig:multi}, 
showing that the (re-scaled) singularity spectrum
of critical wave-functions broadens as $d$ is increased.
In particular, the lower edge $\alpha_-$ of the support of $f(\alpha)$ seems to approach zero as $d$ is 
increased and $f(\alpha)$ seems to approach (even though there is still a substantial difference) the 
infinite dimensional prediction---observed on tree-like lattices--- which corresponds to the strongest possible 
form of multifractality and is represented by the straight line in fig.~\ref{fig:multi}~\cite{evers}.\footnote{We 
were not able to compute the singularity spectrum
in $6$ dimensions due to practical limitations in the system sizes.}
These observations support once again the extreme form of AL criticality in the $d \to \infty$
limit, where the critical states
correspond to an insulator, are described by Poisson statistics, 
and their multifractal spectrum takes its strongest possible form.

\subsubsection{The $d\rightarrow \infty$ limit and the Bethe Lattice}
We have shown above that the $d\rightarrow \infty$ limit is an extremely good starting point 
to analyze AL in finite dimensions. In usual phase transition the mean-field theory corresponding to 
the large $d$-limit is provided by the exact solution on completely connected models. In the case 
of AL instead completely connected lattices do not provide interesting results and the mean-field theory 
is instead believed to correspond to AL on Bethe lattices~\cite{abou}. Our results confirm this expectation: 
the critical values of all observables tend for $d\rightarrow \infty$ to the ones of the localized phase,
$\overline{r}_c = \overline{r}_P$, $q^{typ}_c = 0$, $\beta_c = 0$, i.e. to the same critical behavior obtained 
for AL on Bethe lattices and tree-like structures \cite{infinitedexact,SUSY-tree}. 

\section{Conclusion, physical picture and perspectives} \label{sec:conclusions}

In the last part of this section, we discuss the implications of our results on the
qualitative and quantitative understanding of AL on finite dimensional lattices. 
\begin{itemize}
\item {\bf Anderson localization and rarefied conducting paths.} The fact that 
the $d\rightarrow \infty$ limit provides a very good starting point to  quantitatively describe
AL suggests that the solution of AL on Bethe lattices is a good starting point to 
get a physical picture of AL on finite dimensional lattices. 
Recently the delocalized phase of the Anderson model on tree-like structures (and on related $d\to \infty$ random
matrix models with long-range hopping~\cite{levy})
has attracted a lot of attention~\cite{noi,scardicchio,altshuler,mirlin1,mirlin2,lemarie}. Although it is still debated whether before the AL transition 
there is a non-ergodic delocalised phase or very strong cross-over regime, it is clear that localisation 
is related to the rarefaction of paths over which electrons can travel, as anticipated in~\cite{SUSYdinf,A97}. Some authors advocates that this leads to a bona-fide multi-fractal intermediate non-ergodic but delocalised phase~\cite{A97,scardicchio,altshuler}, others that 
this picture is valid below a certain scale that diverges (extremely fast) approaching the transition \cite{mirlin1,mirlin2,lemarie}. Although we do not see numerical evidences of an intermediate 
non-ergodic delocalised phase for large $d$, the fact that in the scaling variables that govern finite size scaling 
the linear size of the system, $L=N^{1/d}$, enters raised to powers that remain finite for $d\rightarrow \infty$ suggests that (1)~quasi one-dimensional paths are indeed the relevants geometrical objects for AL in high dimensions, (2)~scaling becomes logarithmic in $N$ for  $d\rightarrow \infty$ as found for tree-like 
structures~\cite{levy} and Bethe lattices~\cite{mirlin1,lemarie}. In summary the idea of non-ergodic transport along rarified paths is relevant even in finite dimension even though possibly only on finite but very large length-scales (on larger ones 
transport would be instead described by standard diffusion).  
\item {\bf Expansion around the Bethe lattice.} In usual phase transitions two different expansions have been developed in order to describe the critical properties: one around the upper and another around the lower critical dimension.
Our results clearly indicate that the former is a much better starting point for AL, see for example the comparison 
for the value of $\nu$ in fig.~\ref{fig:nu}. This is certainly a direction for future research
and suggests that a $1/d$ expansion of the NL$\sigma$M (in its replicated or SUSY formulation) 
could provide an excellent and controlled framework for AL.
It would also be interesting to apply the SDRG to higher dimensions (preliminary results in this direction
are already available~\cite{sdrg} up to $d=10$) as well as to implement alternative real-space RG schemes (such as 
the ``resonance RG'' method~\cite{resonance} and the
Wegner flow equation approach~\cite{WF})
introduced for the family of the power-law random banded matrix ensembles, 
which have been shown to be appropriate RG schemes in the
strong disorder limit. It seems that the only unique framework which would be capable to span the whole range from
the infinitely weak disorder regime (in $d=2 + \epsilon$) to the infinitely strong disorder limit
(for $d \to \infty$) is a non-perturbative RG approach~\cite{nprg}. Developping such an RG method for AL
is certainly worth future studies. 

\end{itemize}

In summary, our work sheds new lights on the critical properties of AL, it characterizes the infinite dimensional limit and  stresses its relevance to describe AL even in three dimensions. Our results are also relevant for 
cases in which localization takes place on infinite dimensional spaces, such as for Many Body Localized systems.

\begin{acknowledgments}
We warmly thank V. Dobrosavljevic, Y. Fyodorov, V. Kravtsov and Gabriel Lemari\'e for useful inputs, remarks and discussions. GB acknowledges support from the ERC grant NPRGGLASS and by a grant from the Simons Foundation (\#454935, Giulio Biroli).
\end{acknowledgments}

\end{document}